\documentclass[%
reprint, 
superscriptaddress,
 amsmath,amssymb,
 aps
]{revtex4-2}
\usepackage[utf8]{inputenc}
\usepackage{graphicx}
\usepackage{dcolumn}
\usepackage{bm}
\usepackage{hyperref}
\usepackage[mathlines]{lineno}
\usepackage{amsmath} 

\relax

\begin{document}

\preprint{APS/123-QED}

\title{Taiwan Axion Search Experiment with Haloscope: CD102 Analysis Details}%

\author{Hsin~Chang}\affiliation{Department of Physics, National Central University, Taoyuan City 320317, Taiwan}
\author{Jing-Yang~Chang}\affiliation{Department of Physics, National Central University, Taoyuan City 320317, Taiwan} 
\author{Yi-Chieh~Chang}\affiliation{National Synchrotron Radiation Research Center, Hsinchu 300092, Taiwan} 
\author{Yu-Han~Chang}\affiliation{Department of Physics, National Chung Hsing University, Taichung City 402202, Taiwan}
\author{Yuan-Hann~Chang}\affiliation{Institute of Physics, Academia Sinica, Taipei City 115201, Taiwan}
\affiliation{Center for High Energy and High Field Physics, National Central University, Taoyuan City 320317, Taiwan}
\author{Chien-Han~Chen}\affiliation{Institute of Physics, Academia Sinica, Taipei City 115201, Taiwan} 
\author{Ching-Fang~Chen}\affiliation{Department of Physics, National Central University, Taoyuan City 320317, Taiwan}
\author{Kuan-Yu~Chen}\affiliation{Department of Physics, National Central University, Taoyuan City 320317, Taiwan} 
\author{Yung-Fu~Chen}\affiliation{Department of Physics, National Central University, Taoyuan City 320317, Taiwan} 
\author{Wei-Yuan~Chiang}\affiliation{National Synchrotron Radiation Research Center, Hsinchu 300092, Taiwan}
\author{Wei-Chen~Chien}\affiliation{Department of Physics, National Chung Hsing University, Taichung City 402202, Taiwan}
\author{Hien~Thi~Doan}\affiliation{Institute of Physics, Academia Sinica, Taipei City 115201, Taiwan} 
\author{Wei-Cheng~Hung}\affiliation{Department of Physics, National Central University, Taoyuan City 320317, Taiwan}\affiliation{Institute of Physics, Academia Sinica, Taipei City 115201, Taiwan} 
\author{Watson~Kuo}\affiliation{Department of Physics, National Chung Hsing University, Taichung City 402202, Taiwan} 
\author{Shou-Bai~Lai}\affiliation{Department of Physics, National Central University, Taoyuan City 320317, Taiwan} 
\author{Han-Wen~Liu}\affiliation{Department of Physics, National Central University, Taoyuan City 320317, Taiwan} 
\author{Min-Wei~OuYang}\affiliation{Department of Physics, National Central University, Taoyuan City 320317, Taiwan}
\author{Ping-I~Wu}\affiliation{Department of Physics, National Central University, Taoyuan City 320317, Taiwan} 
\author{Shin-Shan~Yu}\email[Correspondence to: ]{syu@phy.ncu.edu.tw}\affiliation{Department of Physics, National Central University, Taoyuan City 320317, Taiwan}
\affiliation{Center for High Energy and High Field Physics, National Central University, Taoyuan City 320317, Taiwan}

\collaboration{TASEH Collaboration}

\begin{abstract}
This paper presents the analysis of the data acquired during the first 
physics run of the Taiwan Axion Search Experiment with Haloscope (TASEH), 
a search for axions using a microwave cavity at frequencies between 4.70750 
and 4.79815~GHz. The data were collected from October 13, 2021 to November 15,
 2021, and are referred to as the CD102 data. The analysis of the TASEH CD102 
data excludes models with the axion-two-photon coupling 
$\left|g_{a\gamma\gamma}\right|\gtrsim 8.2\times 10^{-14}\,\text{Ge\hspace{-.08em}V}^{-1}$, a factor of eleven above the benchmark 
KSVZ model for the mass range 
$19.4687 < m_a < 19.8436 \,\mu\text{e\hspace{-.08em}V}$. 
\end{abstract}

\maketitle

\section{Introduction} \label{sec:intro}
The axion is a hypothetical particle predicted as a consequence of a  
solution to the strong CP problem~\cite{strongCPI,strongCPII,strongCPIII}, 
i.e. why the CP symmetry is conserved in the strong 
interaction when there is an explicit CP-violating term in the QCD 
Lagrangian. In other words, why is the electric dipole moment 
of the neutron so tiny:  
$\left|d_n\right| < 1.8 \times10^{-26}~e\cdot\mathrm{cm}$ at 90\% 
confidence level (C.L.)~\cite{EDM,PDG}? 
The solution proposed by Peccei and Quinn is to introduce a new global 
Peccei-Quinn U(1)$_\mathrm{PQ}$ symmetry that is spontaneously broken; the 
axion is the pseudo Nambu-Goldstone boson of 
U(1)$_\mathrm{PQ}$~\cite{strongCPI}. 
Axions are abundantly produced during the QCD phase transition in 
the early universe and may constitute the dark matter 
(DM)~\cite{ADDONI,ADDONII,ADDONIII,ADDONIV}. 
In the post-inflationary PQ symmetry breaking scenario, where the PQ symmetry
is broken after inflation, current calculations suggest a mass range of 
${\cal O}(1–100)\,\mu\text{e\hspace{-.08em}V}$ for axions so that the cosmic 
axion density does not exceed the 
observed cold DM density~\cite{QCDCalI,QCDCalII,QCDCalIII,QCDCalIV,QCDCalV,QCDCalVI,QCDCalVII,QCDCalVIII,QCDCalIX,QCDCalX,QCDCalXI,QCDCalXII,QCDCalXIII}. 
Therefore, axions are compelling because they may explain at the same 
time two puzzles that are on scales different by more than thirty orders of 
magnitude. 

Axions could be detected and studied via their two-photon interaction, the
so-called ``inverse Primakoff effect''. For QCD axions, i.e. the axions 
proposed to solve the strong CP problem, the axion-two-photon coupling 
constant $g_{a\gamma\gamma}$ is related to the mass of the axion $m_a$: 
\begin{equation}
 g_{a\gamma\gamma}= \left(\frac{g_{\gamma}\alpha}{\pi \Lambda^2}\right)m_a, 
\label{eq:grelation}
\end{equation}
where $g_{\gamma}$ is a dimensionless model-dependent parameter, $\alpha$ is 
the fine-structure constant, $\Lambda=78\,\text{Me\hspace{-.08em}V}$ is a 
scale parameter that can be derived from the mass and the decay constant of 
the pion and the ratio of the up to down quark masses. 
The numerical values of $g_{\gamma}$ are -0.97 and 0.36 
in the Kim-Shifman-Vainshtein-Zakharov (KSVZ)~\cite{KSVZI,KSVZII} and 
the Dine-Fischler-Srednicki-Zhitnitsky (DFSZ)~\cite{DFSZI,DFSZII} benchmark 
models, respectively.

The detectors with the best sensitivities to axions with a mass of 
$\approx \,\mu\text{e\hspace{-.08em}V}$, as first put forward by 
Sikivie~\cite{SikivieI,SikivieII}, are haloscopes consisting of a microwave 
cavity immersed in a strong static magnetic field and operated at a cryogenic 
temperature. 
Via the two-photon coupling process, an axion in an external magnetic field 
can convert to an equal-energy photon with a frequency $f$ that is set by: 
$hf=E_a=m_a c^2 + \frac{1}{2}m_a v^2$. The converted photons can be 
accumulated in a cavity with the resonant frequency matched with $f$ and 
subsequently be detected by a signal receiver through the readout probe with 
an adequate coupling to the cavity. 
The axion mass is unknown, therefore, the cavity resonator must allow the 
possibility to be tuned through a range of possible axion masses. The Axion 
Dark Matter eXperiment (ADMX), one of the flagship dark matter search 
experiments, had developed and improved the cavity design and readout 
electronics over the years. The results from the previous 
versions of ADMX and the Generation 2 ADMX (ADMX G2) excluded the KSVZ 
benchmark model within the mass range of 
1.9--4.2$\,\mu\text{e\hspace{-.08em}V}$ and the DFSZ benchmark model for the 
mass ranges of 2.66--3.31 and 3.9--4.1$\,\mu\text{e\hspace{-.08em}V}$, 
respectively~\cite{ADMXI,ADMXII,ADMXIII,ADMXIV,ADMXV,ADMXVI,ADMXVII}. 
One of the major goals of ADMX G2 is to search for higher-mass axions in the 
range of 4--40$\,\mu\text{e\hspace{-.08em}V}$ (1--10~GHz), which is also the 
aim of the new haloscope experiments established during the last ten years.  
The Haloscope at Yale Sensitive to Axion Cold dark matter 
(HAYSTAC) had performed searches first for the mass range of 
23.15--24$\,\mu\text{e\hspace{-.08em}V}$~\cite{HAYSTACIII,HAYSTACIV} and later 
at around 17$\,\mu\text{e\hspace{-.08em}V}$~\cite{HAYSTACI}; they excluded 
axions with 
$\left|g_{\gamma}\right|\geq 1.38 \left|g_{\gamma}^\text{KSVZ}\right|$ 
for $m_a=16.96-17.12$ and 
17.14--17.28$\,\mu\text{e\hspace{-.08em}V}$~\cite{HAYSTACI}. The Center 
for Axion and Precision Physics Research (CAPP) constructed 
and ran simultaneously several experiments targeting at 
different frequencies~\cite{CAPPII,CAPPIII,CAPPI}; 
they have pushed the limits towards the KSVZ value within a narrow mass 
region of 10.7126--10.7186$\,\mu\text{e\hspace{-.08em}V}$~\cite{CAPPI}.
The QUest for AXions-$a\gamma$ (QUAX-$a\gamma$) also pushed their limits 
close to the upper bound of the QCD axion-two-photon couplings for 
$m_a\approx43\,\mu\text{e\hspace{-.08em}V}$~\cite{QUAX}.   

This paper presents the analysis details of a search for axions for the mass 
range of 19.4687--19.8436$\,\mu\text{e\hspace{-.08em}V}$, 
from the Taiwan Axion Search Experiment with Haloscope (TASEH). 
The expected axion signal power and signal line shape, the noise power, 
and the signal-to-noise ratio are described in 
Secs.~\ref{sec:introsignal}--\ref{sec:intronoise}. An overview 
of the TASEH experimental setup is presented in Sec.~\ref{sec:taseh}. 
Section~\ref{sec:calibration} gives a brief description of the calibration for 
the whole amplification chain while Sec.~\ref{sec:ana} details the analysis 
procedure. Section~\ref{sec:faxion} presents the analysis of the synthetic 
axion data and Sec.~\ref{sec:sys} discusses the systematic uncertainties that 
may affect the limits on $\left|g_{a\gamma\gamma}\right|$. 
The final results and the conclusion are presented in Sec.~\ref{sec:results} 
and Sec.~\ref{sec:conclusion}, respectively.

\subsection{The expected axion signal power and signal line shape}
\label{sec:introsignal}

The axion-photon conversion signal power extracted from a microwave cavity on 
resonance is given 
by~\cite{AxionFormula,HAYSTACIII}:
\begin{equation}
P_s = \left(g_{a\gamma\gamma}^2\frac{\hbar^3c^3\rho_a}{m_a^2}\right)\times
\left(\omega_c\frac{1}{\mu_0}B_0^2VC_{mnl}Q_L\frac{\beta}{1+\beta}\right),
\label{eq:ps}
\end{equation}
where $\rho_a=0.45\,\text{Ge\hspace{-.08em}V}/\mathrm{cm}^3$ is the local 
dark-matter density. Both $0.45\,\text{Ge\hspace{-.08em}V}/\mathrm{cm}^3$ 
(used by ADMX, HAYSTAC, CAPP, and QUAX) and 
$0.3\,\text{Ge\hspace{-.08em}V}/\mathrm{cm}^3$ (more commonly cited 
by the other direct DM search experiments) are consistent with the recent 
measurements~\cite{Read:2014qva,PDG}. 
The second set of parameters are related to the experimental 
setup and include: the angular resonant frequency of the cavity $\omega_c$, 
the vacuum permeability $\mu_0$, the nominal strength of the external magnetic 
field $B_0$, the volume of the cavity $V$, and the loaded quality factor of 
the cavity \(Q_L=Q_0/(1+\beta)\), where $Q_0$ is the unloaded, intrinsic 
quality factor of the cavity and $\beta$ is the coupling coefficient which 
determines the amount of coupling of the signal to the receiver. The form 
factor $C_{mnl}$ is the normalized overlap of the electric field 
$\vec{\bm{E}}$, for a particular cavity resonant mode, with the external 
magnetic field $\vec{\bm{B}}$:
\begin{equation}
  C_{mnl} = \frac{\left[\int\left( \vec{\bm{B}}\cdot\vec{\bm{E}}_{mnl}\right) d^3\bm{x}\right]^2}{B_0^2V\int E_{mnl}^2 d^3\bm{x}}.
\label{eq:formfactor} 
\end{equation} 
The magnetic field $\vec{\bm{B}}$ in TASEH points mostly along the axial 
direction of the cavity, with a small variation of 
field strength along the radial and axial directions.  
For cylindrical cavities, the largest form factor is from the 
TM$_{010}$ mode. The expected signal power derived from the experimental 
parameters of TASEH (see Table~\ref{tab:tasehbenchmark}) 
is $P_s\simeq 1.4\times10^{-24}$~W for a KSVZ axion with a 
mass of 19.5$\,\mu\text{e\hspace{-.08em}V}$. 

In the direct dark matter search experiments, several assumptions are 
made in order to derive a signal line shape. 
The density and the velocity distributions of DM are related to each other 
through the gravitational potential. The DM in the galactic halo is assumed 
to be virialized. The DM halo density distribution is assumed 
to be spherically symmetric and close to be isothermal, which results in a 
velocity distribution similar to the Maxwell-Boltzmann distribution. The 
distribution of the measured signal frequency can be further derived from the 
velocity distribution after a change of variables and set 
\(hf_a = m_a c^2\). 
For frequency $f\ge f_a$:
\begin{equation}
\mathcal{F}(f, f_a) = \frac{2}{\sqrt{\pi}}\sqrt{f-f_a}\left(\frac{3}{\alpha}\right)^{3/2}
e^{\frac{-3\left(f-f_a\right)}{\alpha}}, 
\label{eq:simplesignal}
\end{equation}
where $\alpha\equiv  f_a \left<v^2\right>/c^2$. Previous axion searches 
typically adopt Eq.~\eqref{eq:simplesignal} when deriving their analysis 
results~\cite{HAYSTACII}. For a Maxwell-Boltzmann velocity 
distribution, the variance $\left<v^2\right>$ and the most probable velocity 
(speed) $v_p$ are related to each other:
$\left<v^2\right>=3v_p^2/2=$(270~km/s)$^2$, where $v_p=220$~km/s is the local 
circular velocity of DM in the galactic rest frame and this value is also 
used by other axion experiments. 

Equation~\eqref{eq:simplesignal} 
is modified if one considers that the relative velocity of the DM halo with 
respect to the Earth is not the same as the DM velocity in the galactic rest 
frame~\cite{SignalLineShapeI}. The velocity distributions shall also be 
truncated so that the DM velocity is not larger than the escape velocity of 
the Milky Way~\cite{Lisanti:2016jxe}. 
Several numerical simulations follow structure formation from the initial DM 
density perturbations to the largest halo today and take into account the 
merger history of the Milky Way, rather than assuming that the Milky Way is 
in a steady state. 
Earlier high-resolution DM-only simulations suggested velocity distributions 
noticeably different from the Maxwellian 
one~\cite{PDG,Lisanti:2016jxe,Green:2017odb}. 
The recent hydrodynamical simulations including baryons, which have a 
non-negligible effect on the DM distribution in the Solar neighborhood, find 
that the velocity distributions are closer to Maxwellian than previously 
thought~\cite{PDG,Green:2017odb}. However, there may still be deviations and 
significant variations depending on the detailed characteristics of the halos.
 By studying the motion of stars that are expected to have the same kinematics 
as the DM, one could determine the DM velocity distribution from observations. 
The data from the Gaia satellite~\cite{GAIA} imply that the local DM halo, 
similar to the local stellar halo, may have a component that is quasi-spherical
 and a component that is radially anisotropic, giving a velocity distribution 
slightly shifted towards higher values with respect to the Maxwellian 
one~\cite{Evans:2018bqy}.  

In order to compare the results of TASEH with those of the former axion 
searches, the analysis presented in this paper uses the axion signal line shape
 from Eq.~\eqref{eq:simplesignal} (see Sec.~\ref{sec:merge}). 
A signal line width $\Delta f_a=m_a\left<v^2\right>/h\simeq$~5~kHz, which is 
much smaller than the TASEH cavity line width $f_a/Q_L\simeq$~240~kHz, is 
assumed. For a signal line shape as described in Eq.~\eqref{eq:simplesignal}, 
a 5-kHz bandwidth includes about 95\% of the distribution. 
Still given the caveats above and a lack of strong evidence for any particular
 choice of the velocity distribution, two different scenarios are considered 
and their results are presented for comparison: (i) without an assumption of 
signal line shape, and (ii) assuming a Gaussian signal line shape with a 
narrower full width at half maximum (FWHM), see Sec.~\ref{sec:results} for more
 details.

\subsection{The expected noise and the signal-to-noise ratio}
\label{sec:intronoise}
Several physics processes can contribute to the total noise and all of them 
can be seen as Johnson thermal noise at some effective temperature, or the 
so-called system noise temperature $T_\text{sys}$. The total noise power in a 
bandwidth $\Delta f$ is then:
\begin{equation}
  P_n = k_B T_\text{sys} \Delta f, 
\end{equation}
where $k_B$ is the Boltzmann constant. 
The system noise temperature $T_\text{sys}$ has three major components: 
\begin{equation}
 T_\text{sys} = \Tilde{T}_\text{mx} + \left(\Tilde{T}_\text{c}-\Tilde{T}_\text{mx}\right)L(\omega) + T_\text{a},
\label{eq:pn}
\end{equation}
where $\omega$ is the angular frequency. 
The last term $T_\text{a}$ is the effective temperature of the 
noise added by the receiver. 
The sum of the first two terms, 
$\Tilde{T}_\text{mx} + \left(\Tilde{T}_\text{c}-\Tilde{T}_\text{mx}\right)L(\omega)$, 
is equivalent to the sum of the noise reflected by the cavity from the 
attenuator anchored to the mixing flange and the noise from the cavity body 
itself. The symbol 
$\Tilde{T}_i=\left(\frac{1}{e^{\left.\hbar\omega\middle/k_BT_i\right.}-1} + \frac{1}{2}\right)\left. \hbar\omega\middle/k_B\right.$ refers to the effective 
temperature due to the blackbody radiation at a physical temperature $T_i$ and 
the quantum noise associated with the zero-point fluctuation of the vacuum. 
The values $T_\text{c}\simeq155$~mK and $T_\text{mx}\simeq27$~mK are the 
physical temperatures of the cavity and of the mixing flange in the dilution 
refrigerator, respectively (see Sec.~\ref{sec:taseh}). The difference 
of the effective temperatures $\Tilde{T}_\text{c}-\Tilde{T}_\text{mx}$ is 
modulated by a Lorentzian function $L(\omega)$. 
The derivation of the first two terms in Eq.~\eqref{eq:pn} can be found in 
Appendix~\ref{sec:cavitynoise}.

Using the operation parameters of TASEH in Table~\ref{tab:tasehbenchmark} and 
the results from the calibration of readout electronics, 
the baseline value of $T_\text{sys}$ for TASEH 
is about 2.0--2.3~K, which gives a noise power of approximately 
$\left(1.4-1.6\right)\times 10^{-19}$~W within the 5-kHz axion signal 
line-width, five orders of magnitude larger than the signal. Nevertheless, what
 matters in the analysis is the signal significance, or the so-called 
signal-to-noise ratio (SNR) using the standard terminology of axion 
experiments, i.e. the ratio of the signal power to the fluctuation in the 
averaged noise power spectrum $\sigma_n$.

According to Dicke's Radiometer Equation~\cite{Dicke}, $\sigma_n$ 
is given by: 
\begin{eqnarray}
 \sigma_n  &=&  \frac{P_n}{\sqrt{N_\text{avg}}}, \nonumber \\
           &=&  \frac{P_n}{\sqrt{t\Delta f}}, \nonumber \\
           &=&  k_B T_\text{sys}\sqrt{\frac{\Delta f}{t}} 
 \label{eq:sigman}
\end{eqnarray}
where $N_\text{avg}$ is the number of noise power spectra used in the 
average; it is related to the data integration time $t$ 
and the resolution bandwidth $\Delta f$.  
Assuming that all the axion signal power falls within $\Delta f$, 
the SNR will therefore be: 
\begin{eqnarray}
   \text{SNR} & = & \frac{P_s}{\sigma_n}, \nonumber \\
              & = & \frac{P_s}{k_B T_\text{sys}}\sqrt{\frac{t}{\Delta f}},
 \label{eq:SNR}
\end{eqnarray}  
Combining Eq.~\eqref{eq:ps} and Eq.~\eqref{eq:SNR},
one could see that the SNR is maximized by an experimental setup with 
a strong magnetic field, a large cavity volume, an efficient cavity 
resonant mode, a receiver with low system noise temperature, and a 
long integration time. 

\section{Experimental Setup}\label{sec:taseh} 
The detector of TASEH is located at the Department of Physics, National 
Central University, Taiwan and housed within a cryogen-free dilution 
refrigerator (DR) from BlueFors. An 8-Tesla superconducting solenoid 
with a bore diameter of 76~mm and a length of 240 mm is integrated with the 
DR. 

While taking data, the cavity sits in the center of the magnet bore 
and is connected to the mixing flange of the DR. 
The cavity, made of oxygen-free high-conductivity (OFHC) copper, has an 
effective volume of 0.234~L and is a cylinder split into two halves along 
the axial direction~\cite{TASEHInstrumentation}. 
The cylindrical cavity has an inner radius of 2.5~cm and a 
height of 12~cm.  In order to maintain a smooth surface, the cavity underwent 
the processes of annealing, polishing, and chemical cleaning. The resonant 
frequency of the TM$_{010}$ mode at the cryogenic temperature 
can be tuned over the range of 
4.667--4.959~GHz via the rotation of an off-axis OFHC copper tuning rod, from 
the position closer to the cavity wall to the position closer to the cavity 
center (i.e. when the vector from the rotation axis to the tuning rod is 
at an angle of $0^\circ$ to $180^\circ$, with respect to the vector from the 
cavity center to the rotation axis). 
The values of the form factor $C_{010}$, as defined in 
Eq.~\eqref{eq:formfactor}, 
 are derived from the magnetic field map provided by 
BlueFors and the cavity electric field distribution simulated with 
 Ansys HFSS (high-frequency structure simulator).  

An output probe, made of a 50-$\Omega$ semi-rigid coaxial cable that is 
soldered to an SMA (SubMiniature version A) connector, is inserted into the 
cavity and its depth is set for 
$\beta\simeq2$; the optimization of the value of $\beta$
is discussed in more detail in Ref.~\cite{TASEHInstrumentation}. 
  The signal from the output probe is directed to an 
impedance-matched amplification chain. The first-stage amplifier is  
a low noise high-electron-mobility transistor (HEMT) amplifier with an 
effective noise temperature of $\approx 2$~K, mounted on the 4~K flange. 
The signal is further amplified at room temperature via a 
three-stage post-amplifier, and down-converted 
and demodulated to in-phase (I) and quadrature (Q) components and digitized 
by an analog-to-digital converter with a sampling rate of 2~MHz. 

The data for the analysis presented in this paper were collected by TASEH 
from October 13, 2021 to November 15, 2021, and are termed as the CD102 data, 
where CD stands for ``cool down''. The CD102 data cover the frequency range of 
4.70750--4.79815~GHz. In this paper, most of the frequencies in unit of GHz are
 quoted with five decimal places as the absolute accuracy of frequency is 
$\approx 10$~kHz. It shall be noted that the frequency resolution is 1~kHz.  
The temperature of the cavity stayed at $T_\text{c}\simeq155$~mK, higher 
with respect to the mixing flange temperature $T_{\rm mx}\simeq27$~mK; 
it is believed that the cavity had an unexpected thermal contact with the 
radiation shield in the DR. 
See Table~\ref{tab:tasehbenchmark} for the benchmark experimental parameters 
that can be used to estimate the sensitivity of the CD102 data, including the 
form factor $C_{010}$, the intrinsic, unloaded quality factor $Q_0$ at 
the cryogenic temperature, the number of resonant-frequency steps 
$N_\text{step}$, and the frequency difference between the steps $\Delta f_s$.  
Each resonant-frequency step is denoted as a ``scan'' and the data integration
 time was about 32-42 minutes. The integration time was determined based on 
the target $\left|g_{a\gamma\gamma}\right|$ limits and the experimental 
parameters in Table~\ref{tab:tasehbenchmark}; the variation of the integration 
time aimed to remove the frequency-dependence in the 
$\left|g_{a\gamma\gamma}\right|$ limits caused by frequency dependence of the 
added noise $T_\text{a}$. 

A more detailed description of the TASEH detector, the operation of the 
data run, and the calibration of the gain and added noise temperature of the 
whole amplification chain can be found in Ref.~\cite{TASEHInstrumentation}. 

\begin{table}
\caption{The benchmark experimental parameters for estimating the sensitivity 
of the CD102 data. The definitions of the parameters can be found in 
Sec.~\ref{sec:intro} and Sec.~\ref{sec:taseh}. 
More details regarding the determination and the measurements of 
some of the parameters may be found in Ref.~\cite{TASEHInstrumentation}.} 
\label{tab:tasehbenchmark}
\begin{center}
\begin{tabular}{cr}
\hline\hline
 $f_\mathrm{lo}$ & 4.70750~GHz\\
 $f_\mathrm{hi}$ & 4.79815~GHz \\
 $N_\text{step}$ & 837 \\
 $\Delta f_\text{s}$ & 95 -- 115 kHz \\
 $B_0$  & 8 Tesla \\
 $V$ & 0.234 L \\ 
 $C_{010}$ & 0.60 -- 0.61 \\
 $Q_0$ & 58000 -- 65000 \\
 $\beta$ & 1.9 -- 2.3 \\
 $T_{\rm mx}$ & 27--28~mK\\
 $T_\mathrm{c}$ & 155~mK \\
 $T_\text{a}$ & 1.9--2.2~K \\
 $\Delta f_a$ & 5 kHz \\
\hline\hline
\end{tabular}
\end{center}
\end{table}

\section{Calibration} \label{sec:hemtcalibration}
\label{sec:calibration}
The noise is one of the most important parameters for the axion searches. 
Therefore, calibration for the amplification chain is a 
crucial part in the operation of TASEH. In order to perform a calibration, 
the HEMT is connected to a heat source (a 50-$\Omega$ resistor) instead of 
the cavity; 
various values of input currents are sent to the source to change its 
temperature monitored by a thermometer. The power from the source 
is delivered following the same transmission line as that in the CD102 
run. 
The output power is fitted to a first-order polynomial, as a function of 
the source temperature, to extract the gain and added noise for the 
amplification chain. More details of the 
procedure can be found in Ref.~\cite{TASEHInstrumentation}. 

\begin{figure} [htbp]
  \centering
  \includegraphics[width=8.6cm]{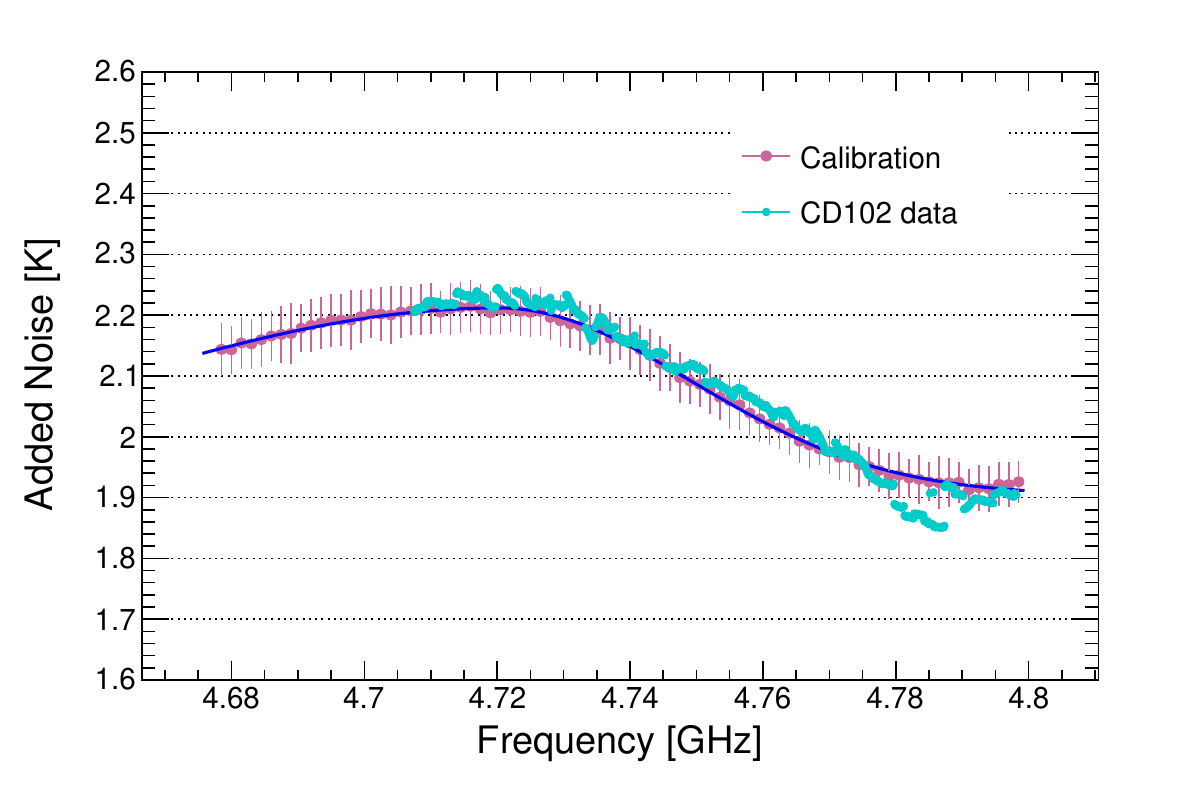}
  \caption{The average added noise obtained from the calibration (pink points)
 and the noise estimated from the CD102 data (light blue points) as a 
function of frequency. The error bars on the pink points are the RMS 
of $T_\text{a}$, as computed from the 19 measurements for each frequency 
in the calibration. 
The blue curve is obtained after performing a fit to 
the pink points and is used to estimate $T_\text{a}$ at the corresponding 
frequency.}
  \label{fig:hemtcalvsf}
\end{figure}

The calibration was carried out before, during, and after the data taking, 
which showed that the performance of the system was stable over time. The 
average of the added noise $T_\text{a}$ over 19 measurements has the lowest 
value of 1.9~K at the frequency of 4.8~GHz and the highest value of 
2.2~K at 4.72~GHz, as presented in Fig.~\ref{fig:hemtcalvsf}. 
The error bars are the RMS of $T_\text{a}$ and the largest RMS is used to 
calculate the systematic uncertainty for the limits on 
$\left|g_{a\gamma\gamma}\right|$. The light blue points in 
Fig.~\ref{fig:hemtcalvsf} are the noise estimated from the CD102 data by 
removing the gain and subtracting the contribution from the cavity noise, 
assuming 
that the presence of a narrow signal in the data would have no effect on the 
estimation. A good agreement between the results from the calibration  
and the ones estimated from the CD102 data is shown. The biggest 
difference is 0.076~K in the frequency range during which the data were 
recorded after an earthquake. The source of the difference is not understood, 
therefore, the difference is quoted as a systematic uncertainty together 
with the RMS of the noise.

\section{Analysis Procedure} \label{sec:ana}
The goal of TASEH is to find the axion signal hidden in the noise. In 
order to achieve this, the analysis procedure includes the following steps:
    \begin{enumerate}
        \item Perform fast Fourier transform (FFT) on the 
IQ time series data to obtain the frequency-domain power spectrum.
        \item Apply the Savitzky-Golay (SG) filter to remove the structure 
of the background in the frequency-domain power spectrum.
        \item Combine all the spectra from different frequency scans with 
the weighting algorithm.
        \item Merge bins in the combined spectrum to maximize the SNR. 
       \item Rescan the frequency regions with candidates and set limits on 
      the axion-two-photon coupling $\left|g_{a\gamma\gamma}\right|$ if no 
candidates were found.
    \end{enumerate}

    The analysis follows the procedure similar to that 
developed by the HAYSTAC experiment~\cite{HAYSTACII}. The important points  
and formulas for each step are highlighted below as a reminder 
for the convenience of readers. Note there are a few  
small differences between the HAYSTAC analysis and the one presented here. 
In this paper, the uncertainties are considered to be uncorrelated between 
different frequency bins while Ref.~\cite{HAYSTACII} takes into account 
the correlation. The frequency-domain spectra processed by each intermediate 
step are shown. The central results of the $\left|g_{a\gamma\gamma}\right|$ 
limits assume the signal line shape described by Eq.~\eqref{eq:simplesignal} 
as in Ref.~\cite{HAYSTACII}. In addition, the limits without an assumption of 
signal line shape and the limits assuming a 
Gaussian signal with a narrower FWHM are 
shown for comparison in Sec.~\ref{sec:results}. 
As a sanity check, the data are analyzed by two 
independent groups and their results are consistent with each other.

\subsection{Fast Fourier transform}
\label{sec:FFT}
The in-phase $I(t)$ and quadrature $Q(t)$ components of the time-domain 
data were sampled and saved in the TDMS 
(Technical Data Management Streaming) files - a 
binary format developed by National Instruments. 
The FFT is performed to convert the data into 
frequency-domain power spectrum in which the power is calculated 
using the following equation:

\begin{equation}
\label{eq:4.1}
    \text{Power} = \frac{|\text{FFT}(I+i \cdot Q)|^{2}}{N \cdot 2R},
\end{equation}
where $N$ is the number of data points ($N  = 2000$ in the TASEH 
CD102 data), and $R$ is the input resistance of the signal analyzer 
(50~$\Omega$).
The FFT is done for every one-millisecond subspectrum data. The integration 
time for each frequency scan was about 32-42 minutes, which resulted 
in 1920000 to 2520000 subspectra; an average over these subspectra gives 
the averaged frequency-domain power spectrum for each scan. 
The frequency span in the spectrum from each resonant-frequency scan is 
2~MHz while the resolution is 1~kHz. In order to avoid the aliasing 
effect, a band-pass filter was applied in the data acquisition, 
giving a frequency span of 1.6~MHz (1600 frequency bins) that can be used 
for the analysis.

\subsection{Remove the structure of the background} 
\begin{figure*} [htbp]
  \centering
  \includegraphics[width=6.5cm]{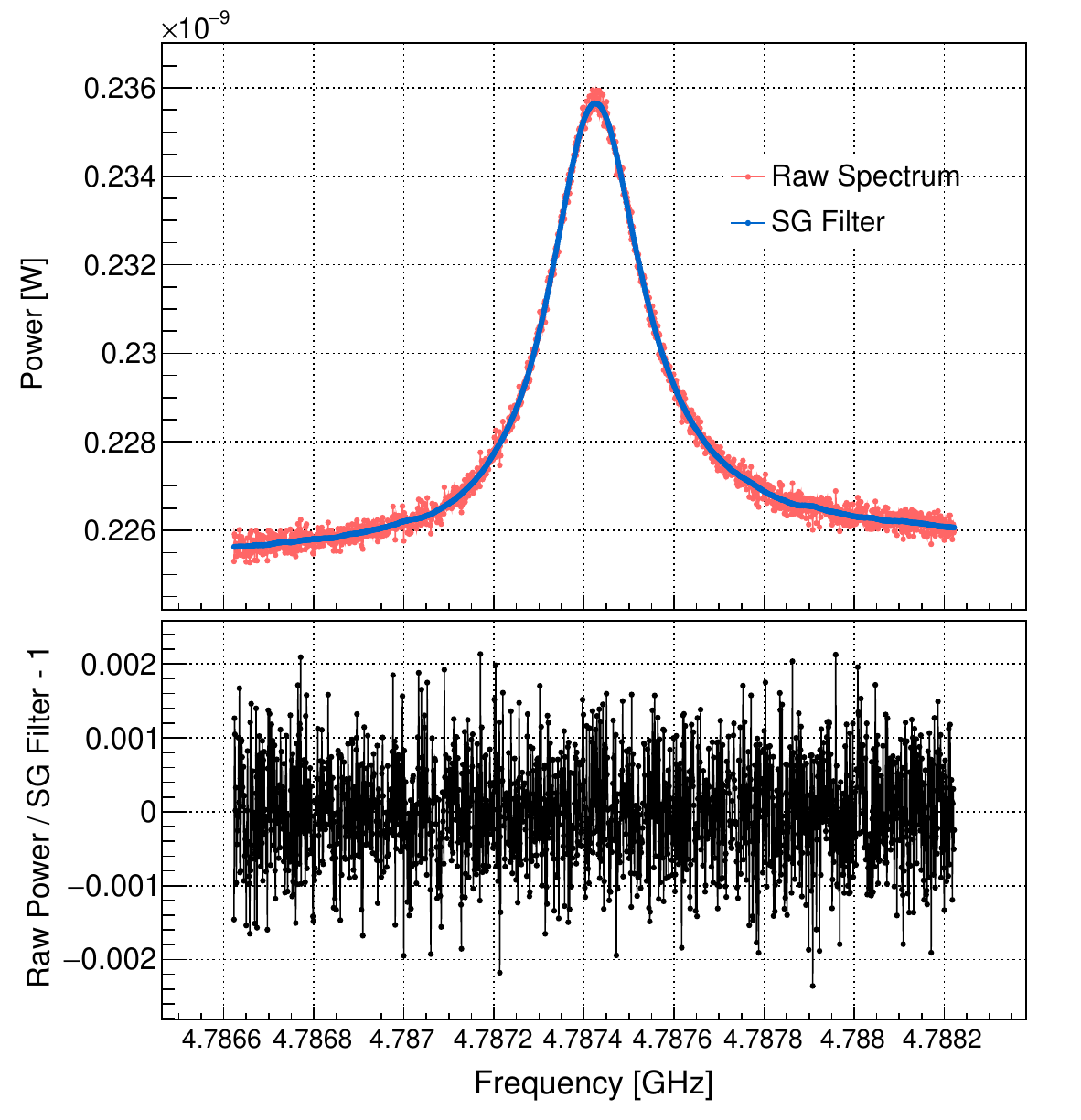}
  \includegraphics[width=6.5cm]{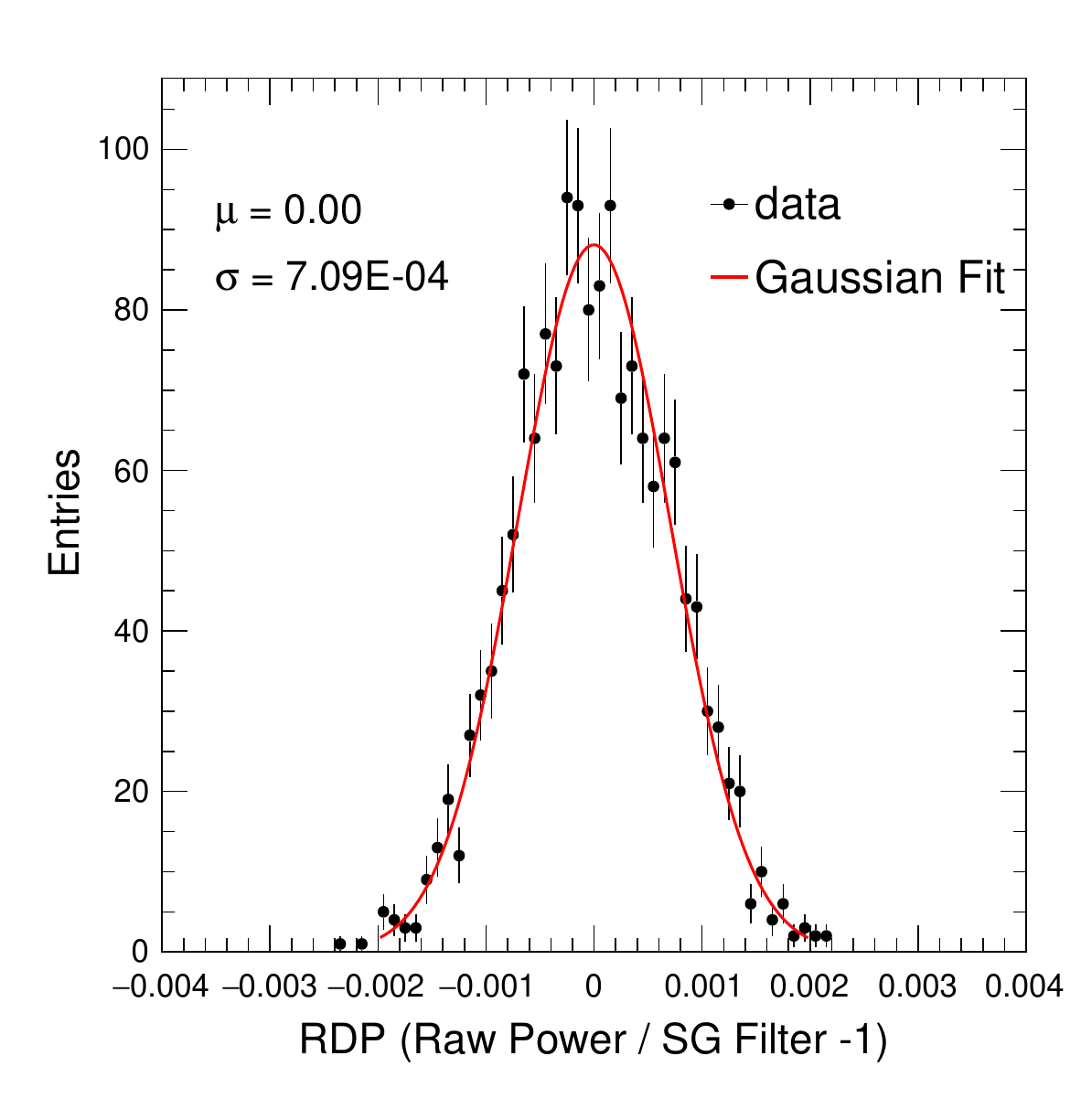}
  \caption{Upper left panel: The raw averaged power spectrum (red points) and 
the output of the SG filter (blue curve) of one scan. Lower left panel: The 
normalized spectrum,  derived by taking the ratio of the raw spectrum to the 
SG filter and subtracting unity from the ratio. Right plot: Histogram of the 
normalized spectrum (lower panel in left plot) with a Gaussian 
fit; there are 1600 entries in total (from the 1600 frequency bins). 
The fitted mean and standard deviation are shown to be consistent with the 
prediction when the axion signal is not present.}
  \label{fig:raw_sg_power}
\end{figure*}

In the absence of the axion signal, the output data spectrum is simply the 
noise from the cavity and the amplification chain. If axions are present 
in the cavity, the signal will be buried in the noise because the 
signal power is very weak. Therefore, the structure of the raw averaged 
output power spectrum, as shown in the upper left panel of 
Fig.~\ref{fig:raw_sg_power}, is dominated 
by the noise of the system and an explanation for the structure can be found 
in Appendix~\ref{sec:cavitynoise}. The SG 
filter~\cite{SGFilter}, a digital filter that can smooth data without 
distorting the signal tendency, is applied to remove the structure of the  
background. The SG filter is performed on the averaged spectrum of each 
frequency scan by fitting adjacent points of successive sub-sets of data with 
an $n^\text{th}$-order polynomial. The result depends on two parameters: 
the number of 
data points used for fitting, the so-called window width, and the order of 
the polynomial. If the window is too wide, the filter will not remove small 
structures, and if it is too narrow, it may kill the signal. 
A window width of 201 frequency bins and a 4$^\text{th}$-order polynomial 
were first chosen during the data taking, by 
requiring the ratio of the raw data to the filter 
output consistent with unity.  
The SG-filter parameters 
are also cross-checked using 10000 pseudo-experiments that include simulations 
of the noise spectrum and an axion signal with 
$\left|g_{a\gamma\gamma}\right|\approx 10 \left|g_{a\gamma\gamma}^\text{KSVZ}\right|$; the measured signal power 
is found to be consistent with the injected one within 1\%.

The SG-filter output can be considered as the averaged noise power. 
The raw averaged power spectrum is divided by the output of the SG filter, 
then unity is subtracted from the ratio to get the dimensionless 
normalized spectrum (lower left panel of Fig.~\ref{fig:raw_sg_power}). 
The relative deviation of power (RDP) in the normalized spectrum (and also in 
the spectra processed with rescaling, combining, and merging afterwards) are 
denoted by the symbol $\delta$. The values of 
RDPs can be zero, positive, or negative. 
In the absence of the axion signal, the RDPs in the normalized spectrum are 
samples drawn from a Gaussian 
distribution with a zero mean and a standard deviation of 
$\left.1\middle/\sqrt{N_\text{spectra}}\right.$, where $N_\text{spectra}$ is 
the number of subspectra used to compute the average (see Sec.~\ref{sec:FFT} 
and the right panel of Fig.~\ref{fig:raw_sg_power}). 

The normalized spectrum from each scan is further rescaled 
 with the following formula:
\begin{equation}
  \label{eq:respower_eqn}
  \delta_{ij}^\text{res} = R_{ij}\delta_{ij}^\text{norm},
\end{equation}
and the standard deviation of each bin is:
\begin{equation}
  \label{eq:ressigma_eqn}
  \sigma_{ij}^\text{res} = R_{ij}\sigma_{i}^\text{norm},
\end{equation}
where 
 \begin{equation}
 R_{ij} = \frac{k_{B}T_\text{sys} \Delta f_\text{bin} }{P_{ij}^\text{KSVZ} h_{ij}}, 
 \label{eq:Rratio}
 \end{equation}
and 
 \begin{equation}
 h_{ij} = \frac{1}{1 + 4Q_{Li}^{2}(\left.f_{ij}\middle/f_{ci}\right.-1)^2}. 
 \label{eq:Lorentz}
 \end{equation}
The notations $\delta_{ij}^\text{norm}$ ($\delta_{ij}^\text{res}$) and 
$\sigma_{i}^\text{norm}$ ($\sigma_{ij}^\text{res}$) are the 
RDP and the standard deviation of the $j^\text{th}$ frequency bin in 
the normalized (rescaled) spectrum from the 
$i^\text{th}$ resonant-frequency scan. 
The value of $\sigma_{i}^\text{norm}$ is derived from the spread of the 
RDPs over the 1600 frequency bins for the $i^\text{th}$ scan 
(see an example in the right panel of Fig.~\ref{fig:raw_sg_power}). 
The factor $R_{ij}$ is the ratio of 
the system noise power to the expected signal power of the KSVZ axion 
$P_{ij}^\text{KSVZ}$, with the Lorentzian cavity response $h_{ij}$ 
taken into account. 
The system-noise temperature $T_\text{sys}$ in Eq.~\eqref{eq:Rratio} is 
calculated following Eq.~\eqref{eq:pn},
 where the frequency dependence of the added-noise temperature $T_\text{a}$ is 
obtained from the fitting function in Fig.~\ref{fig:hemtcalvsf}. 
The symbol $\Delta f_\text{bin}$ is the bin width of spectrum (1~kHz). 
The factor $h_{ij}$ describes the Lorentzian response of the cavity, 
which depends on the loaded quality factor $Q_{Li}$ and the 
difference between the frequency $f_{ij}$ in bin $j$ and the resonant 
frequency $f_{ci}$. 
If a signal appears in a certain frequency bin $j$, its expected power 
will vary depending on the bin position due to the cavity's 
Lorentzian response. The rescaling will take into account this effect. 
The procedure of the normalization and the rescaling also ensures that a 
KSVZ axion signal will have a rescaled RDP $\delta_{ij}^\text{res}$ 
that is approximately equal to unity, if the signal power is distributed 
in only one frequency bin. 

\subsection{Combine the spectra with the weighting algorithm} 
\label{sec:weighting_algorithm}
During the data taking, the resonant frequency of the cavity was  
adjusted by the tuning bar to scan a large range of frequencies.  
Therefore, the spectra of all the scans need to be combined to create one 
big spectrum. 
The purpose of the weighting algorithm is to add the spectra from different 
resonant-frequency scans,
 particularly for the frequency bins that appear in multiple spectra. 
Note that the uncertainty of the averaged power at the overlapped region is 
reduced due to the combination.  
The weight is defined below: 
\begin{equation}
    \label{eq:weight}
    {w_{ijn}} = \frac{\Gamma_{ijn}}{(\sigma_{ij}^\text{res})^{2}}.
\end{equation}
Here, the symbol $\Gamma_{ijn}=1$ if the $j^\text{th}$ frequency bin in the 
$i^\text{th}$ rescaled spectrum correspond to the same frequency in 
the $n^\text{th}$ bin of the combined spectrum; otherwise, $\Gamma_{ijn}=0$.

\begin{figure}[hbt!]
    \centering
    \includegraphics[width=8.6cm]{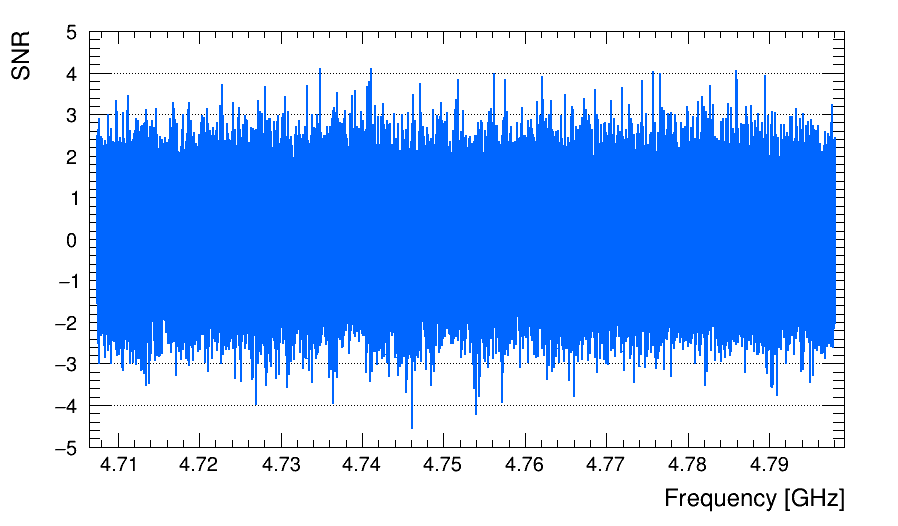}
    \caption{The signal-to-noise ratio (SNR) calculated using 
Eq.\eqref{eq:comb_snr} of the combined spectrum. }
    \label{fig:SNR_comb}
\end{figure}

The RDP $\delta^\text{com}_{n}$ and the standard deviation 
$\sigma^\text{com}_{n}$ of the $n^\text{th}$ bin in the combined spectrum are 
calculated using Eq.~\eqref{eq:comb_power} and Eq.~\eqref{eq:comb_sigma}, 
respectively. The notation SNR$^\text{com}_{n}$ is the ratio of 
$\delta^\text{com}_{n}$ to 
$\sigma^\text{com}_{n}$ as given in Eq.~\eqref{eq:comb_snr}. 
Figure~\ref{fig:SNR_comb} shows the SNR of the combined spectrum. 

\begin{equation}
    \label{eq:comb_power}
    \delta_{n}^\text{com} = \frac{\sum\limits_{i}\sum\limits_{j}\left(\delta_{ij}^\text{res} \cdot {w_{ijn}}\right)}{\sum\limits_{i}\sum\limits_{j} {w_{ijn}}},
\end{equation}

\begin{equation}
    \label{eq:comb_sigma}
    \sigma_{n}^\text{com} = \frac{ \sqrt{\sum\limits_{i}\sum\limits_{j}(\sigma_{ij}^\text{res} \cdot {w_{ijn}})^2}}{\sum\limits_{i}\sum\limits_{j} {w_{ijn}}},
\end{equation}
\begin{equation}
    \label{eq:comb_snr}
    \text{SNR}_{n}^\text{com} = \frac{\delta^\text{com}_{n}}{\sigma^\text{com}_{n}}= \frac{\sum\limits_{i}\sum\limits_{j}\left(\delta_{ij}^\text{res} \cdot {w_{ijn}}\right)}{ \sqrt{\sum\limits_{i}\sum\limits_{j}(\sigma_{ij}^\text{res} \cdot {w_{ijn}})^2}}.
\end{equation} 
The summations over $i$ run from 1 to 837 (steps) while the summations 
over $j$ run from 1 to 1600 (bins).  
For each bin $n$ in the combined spectrum, there are $m_n$ non-vanishing 
contributions to the sums above. In general, the value of $m_n$ is 14--16.

\subsection{Merge bins}
\label{sec:merge}

The expected axion bandwidth is about 5~kHz at the frequency of $\approx5$~GHz.
 In this paper, the interested frequency range is 4.70750 -- 4.79815~GHz and 
the bin width is 1~kHz. Therefore, in order to maximize the SNR, a running 
window of five consecutive bins in the combined spectrum is applied and the 
five bins within each window are merged to construct a final spectrum.  
The purpose of using a running window is to avoid the signal power broken 
into different neighboring bins of the merged spectrum. 
The number of bins for merging is studied by injecting 
simulated axion signals on top of the CD102 data and optimized based 
on the SNR. 

Due to the nonuniform distribution of the axion signal 
[Eq.~\eqref{eq:simplesignal}],
the contributing bins need to be rescaled to have the same RDP, of which the 
standard deviation is used to define the maximum likelihood (ML)
weight for merging. The rescaling is performed by dividing the values of 
$\delta^\text{com}_{g+k-1}$ and $\sigma^\text{com}_{g+k-1}$ in the combined 
spectrum with an integral of the signal line shape $L_{k}$:

\begin{equation}
  \label{eq:Lq_integral}
  L_{k} = \int_{f_a +\delta f_m + (k-1)\Delta f_\text{bin}}^{f_a +\delta f_m + k\Delta f_\text{bin}} \mathcal{F}(f,f_a) \,df,
\end{equation}
where the variable $g$ is the index for the frequency bins in 
the final merged spectrum and 
$k$ is the index within the group of bins for 
merging. The index $g$ runs from 1 to $N-M+1$, where 
the number $N$ is the total number of bins in 
the combined spectrum and $M=5$ is the number of 
merged bin in this analysis. 
The frequency $f_a=\left.m_a c^2\middle/h\right.$ is the axion
frequency, and $\delta f_m$ is the misalignment between $f_a$ and the lower
boundary of the $g^\text{th}$ bin in the merged spectrum.
The function $\mathcal{F}(f,f_a)$ has been defined in 
Eq.~\eqref{eq:simplesignal}.
In order to get a misalignment-independent line shape, instead of using an
$L_{k}$ that depends on the frequency $f_a$ and  $\delta f_m$, the average 
($\bar{L}_{k}$) of $L_{k}$ over the ranges of $f_a$ and $\delta f_m$ is 
used. 
Note that the relative variation of $f_a$ is at most 
90~MHz/5~GHz$\approx2\%$ and the line shape of $\mathcal{F}(f,f_a)$ 
can be considered constant for the full range of the operational 
frequency. Therefore, 
the value of $L_{k}$ has only weak dependence on $f_a$. 
In the analysis presented here, 
$\bar{L}_{k} = 0.23, 0.33, 0.21, 0.11, 0.06$ for $k = 1, ... 5$, respectively.
The effect of the misalignment on the $\left|g_{a\gamma\gamma}\right|$ limits 
is quoted as a part of the systematic uncertainty using the same method as 
described in the HAYSTAC paper~\cite{HAYSTACII}, see Sec.~\ref{sec:sys}.

The rescaled RDP $\delta^\text{rs}_{g+k-1}$ and
standard deviation $\sigma^\text{rs}_{g+k-1}$ are calculated:
\begin{equation}
  \label{eq:rescaled_delta_sigma_com}
  \begin{split}
  \delta^\text{rs}_{g+k-1} = \frac{\delta^\text{com}_{g+k-1}}{\bar{L}_{k}},\\
  \sigma^\text{rs}_{g+k-1} = \frac{\sigma^\text{com}_{g+k-1}}{\bar{L}_{k}}.
  \end{split}
\end{equation}
After this rescaling procedure, a KSVZ axion signal is expected to have an 
RDP equal to unity for each of the five bins. The ML weight is defined as: 
\begin{equation}
    \label{eq:merge_weight}
    w_{gk} = \frac{1}{(\sigma_{g+k-1}^\text{rs})^{2}} = \frac{\bar{L}_{k}^{2}}{(\sigma_{g+k-1}^\text{com})^{2}}.
\end{equation}

The RDP, the standard deviation, and the SNR of the merged spectrum are:

\begin{equation}
    \delta_{g}^\text{merged} = \frac{ \sum\limits_{k = 1}^{M}\left(\delta_{g+k-1}^\text{rs} \cdot {w_{gk}}\right)}{\sum\limits_{k = 1}^{M} {w_{gk}}} = \frac{\sum\limits_{k = 1}^{M}\frac{\delta_{g+k-1}^\text{com}}{\bar{L}_{k}} \cdot \left(\frac{\bar{L}_{k}}{\sigma_{g+k-1}^\text{com}}\right)^2} {\sum\limits_{k = 1}^{M}\left(\frac{\bar{L}_{k}}{\sigma_{g+k-1}^\text{com}}\right)^2},
    \label{eq:merged_power}
\end{equation}

\begin{eqnarray}
  \sigma_{g}^\text{merged} & =  & \frac{ \sqrt{\sum\limits_{k = 1}^{M} \left(\sigma_{g+k-1}^\text{rs} \cdot {w_{gk}}\right)^2}}{\sum\limits_{k = 1}^{M} {w_{gk}}} = \frac{\sqrt{\sum\limits_{k = 1}^{M} \left(\frac{\bar{L}_{k}}{\sigma_{g+k-1}^\text{com}}\right)^2}}{\sum\limits_{k = 1}^{M} \left(\frac{\bar{L}_{k}}{\sigma_{g+k-1}^\text{com}}\right)^2}  \nonumber \\
    & = & \frac{1}{\sqrt{\sum\limits_{k = 1}^{M} \left(\frac{\bar{L}_{k}}{\sigma_{g+k-1}^\text{com}}\right)^2}}
    \label{eq:merged_sigma}
\end{eqnarray}

\begin{equation}
    \label{eq:merged_snr}
    \text{SNR}_{g}^\text{merged} = \frac{\delta^\text{merged}_{g}}{\sigma^\text{merged}_{g}} = \frac{\sum\limits_{k = 1}^{M}\frac{\delta_{g+k-1}^\text{com}}{\bar{L}_{k}} \cdot \left(\frac{\bar{L}_{k}}{\sigma_{g+k-1}^\text{com}}\right)^2}{\sqrt{\sum\limits_{k = 1}^{M} \left(\frac{\bar{L}_{k}}{\sigma_{g+k-1}^\text{com}}\right)^2}}
\end{equation}

\subsection{Rescan and set limits on $\left|g_{a\gamma\gamma}\right|$} 
\begin{figure}[hbt!]
    \centering
    \includegraphics[width=8.6cm]{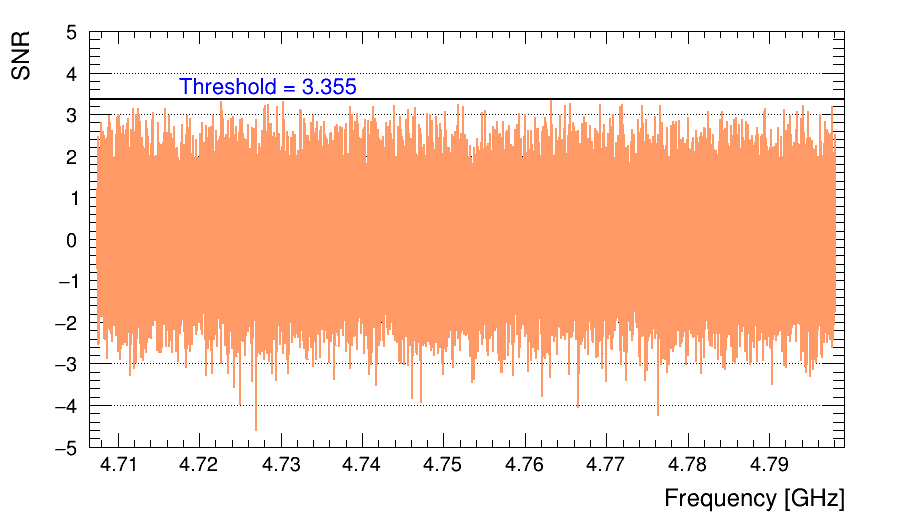}
    \caption{The signal-to-noise ratio (SNR) calculated using Eq.~\eqref{eq:merged_snr} for the merged spectrum including data from both the original 
scans and the rescans. No candidate exceeds the threshold of 
$3.355\sigma$ (solid-black horizontal line). }
    \label{fig:SNR_merged}
\end{figure}

Before the collection of the CD102 data, a 5$\sigma$ SNR target was chosen, 
which corresponds to a candidate threshold of 3.355$\sigma$ at 95\% C.L..
 After the merging as described in Sec.~\ref{sec:merge}, if there were 
any potential signal with an SNR larger than 
3.355, a rescan would be proceeded to check if it were a real signal 
or a statistical fluctuation. 
The procedure of the CD102 data taking was to perform a rescan after 
covering every 10~MHz; the rescan was done by adjusting the tuning rod of the 
cavity so to match the resonant frequency to the frequency of the candidate. 
In total, 22 candidates with an SNR greater than 3.355 were found. 
Among them, 20 candidates were from the fluctuations because they were gone 
after a few rescans. The remaining two candidates, 
in the frequency ranges of 4.71017 -- 4.71019~GHz and 4.74730 -- 4.74738~GHz, 
are excluded from consideration of axion signal candidates due to the 
following reasons. 
The signal in the second frequency range 
was detected via a portable antenna outside the DR and found 
to come from the instrument control computer in the laboratory, while the 
signal in the first frequency range 
was not detected outside the DR but 
still present after turning off the external magnetic field. 
No limits are placed for the two frequency ranges above.  
More details can be found in the 
TASEH instrumentation paper~\cite{TASEHInstrumentation}. 
Figure~\ref{fig:SNR_merged} shows the SNR of the merged spectrum after 
including data from both the original scans and the rescans. 

Since no candidates were found after the rescan, an upper limit on 
the signal power $P_s$ is derived by setting $P_s$ equal to 
$5\sigma_{g}^\text{merged}\times P_{g}^\text{KSVZ}$, where 
$\sigma_{g}^\text{merged}$ is the standard deviation 
and $P_{g}^\text{KSVZ}$ is the expected signal power for the KSVZ axion 
for a certain frequency bin $g$ in the merged spectrum. 
Then, the 95\% C.L. limits on the axion-two-photon coupling 
$\left|g_{a\gamma\gamma}\right|$ could be derived according to 
Eq.~\eqref{eq:ps}. 
See Sec.~\ref{sec:results} for the final limits including the systematic 
uncertainties.

\section{Analysis of the Synthetic Axion Data}\label{sec:faxion}
After TASEH finished collecting the CD102 data on November 15, 2021, 
the synthetic axion signals were injected into the cavity and read out via the 
same transmission line and amplification chain. The procedure 
to generate axion-like signals is summarized in 
Ref.~\cite{TASEHInstrumentation}. 
A test with synthetic axion signals could be used to verify the procedures of 
data acquisition and physics analysis. The synthetic axion signals 
have a wider width (8~kHz) and longer tails compared to the line shape 
described by Eq.~\eqref{eq:simplesignal}. 
The expected SNR of the frequency bin with maximum power ($\approx 11\%$ of 
the total signal power), 
 at 4.70897~GHz, was set to $\approx 3.35$. 
The total signal power injected 
corresponds to $\left|g_{a\gamma\gamma}\right|\approx 20 \left|g_{a\gamma\gamma}^\text{KSVZ}\right|$. 

\begin{figure}[htbp]                                                                                                  
    \centering                                                                                                                       
    \includegraphics[width=8.6cm]{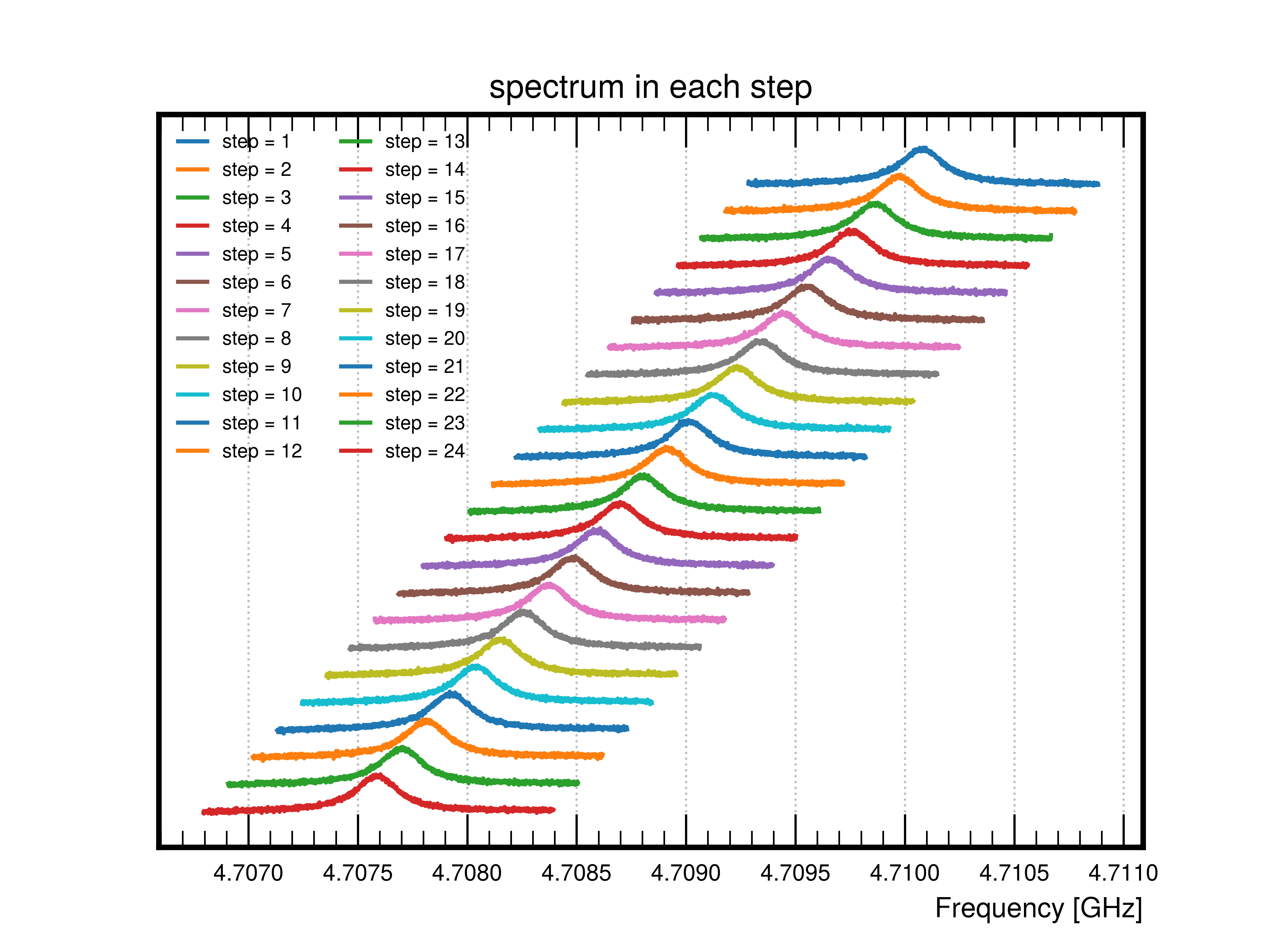}
 \caption{The raw output power spectra, before applying the 
 SG filter, from the 24 frequency steps of the synthetic axion 
data. In order to show the spectra clearly, the spectra are shifted 
with respect to each other with an arbitrary offset in the vertical scale.}                
\label{fig:faxionstep}                                                                                                            
\end{figure}

\begin{figure}[htbp]                                                                                                  
    \centering                                                                                                                       
    \includegraphics[width=8.6cm]{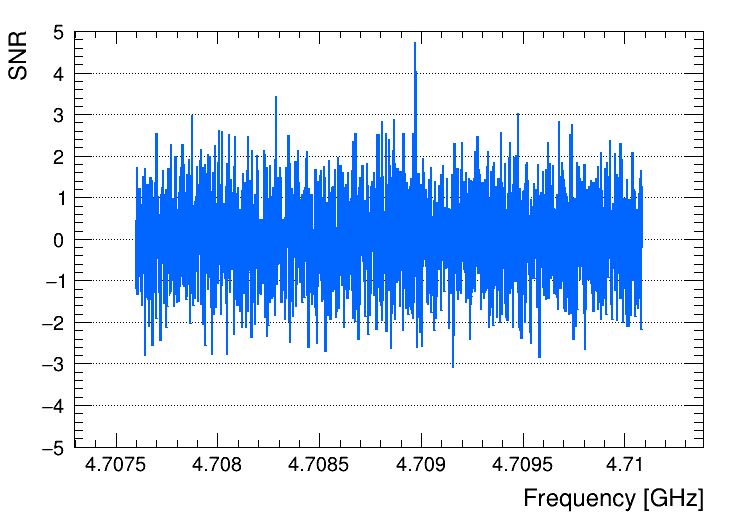}
    \includegraphics[width=8.6cm]{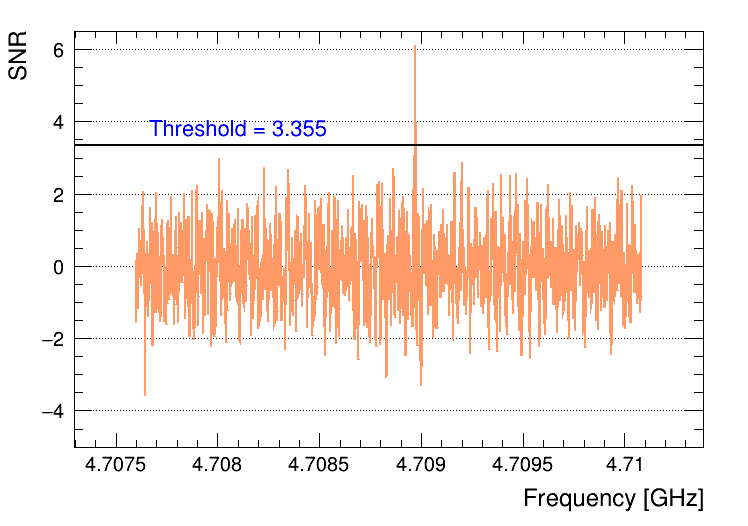}
    \caption{The signal-to-noise ratio, from the synthetic axion data, 
after combining the spectra with overlapping frequencies from different 
scans (upper) and after merging the RDP measured in five neighboring 
frequency bins (lower). 
The procedure and the weights for combination and merging are summarized in 
Sec.~\ref{sec:weighting_algorithm} and Sec.~\ref{sec:merge}, respectively.}               
\label{fig:faxioncombinemerge} 
\end{figure}

The same analysis procedure as described in Sec.~\ref{sec:ana} is applied 
to the data with synthetic axion signals. 
Figure~\ref{fig:faxionstep} presents the individual raw power spectra in 
the 24 frequency scans. Before combining 
the 24 spectra, the SNR of the maximum-power bin from the scan with a resonant 
frequency closest to the injected signal is measured to be 
3.58. 
After the combination of the spectra and the merging of five frequency 
bins, the SNRs of the maximum-power bin increase to 4.74 and 6.12, 
respectively. Figure~\ref{fig:faxioncombinemerge} presents 
the SNR after the combination and the merging, respectively.    
In order to 
validate the results of the SNRs, the analysis procedure is also applied  
to the simulated spectra that include both noise and a signal with the 
same power and the same line shape as those of the injected synthetic axions. 
The SNRs obtained with 200 simulations, before 
the combining, after the combining, and after the merging are 
$3.6\pm 0.5$, $4.5\pm0.6$, and $6.9\pm0.8$, respectively, 
which are 
consistent with the results from the synthetic axion data.  
The consistency of the SNRs demonstrates 
the capability of the TASEH apparatus and the analysis procedure to discover 
an axion signal with 
$\left|g_{a\gamma\gamma}\right|\approx {\cal O}\left(10\left|g_{a\gamma\gamma}^\text{KSVZ}\right|\right)$.

\section{Systematic Uncertainties} \label{sec:sys}
\begin{figure} [htbp]
  \centering
  \includegraphics[width=8.6cm]{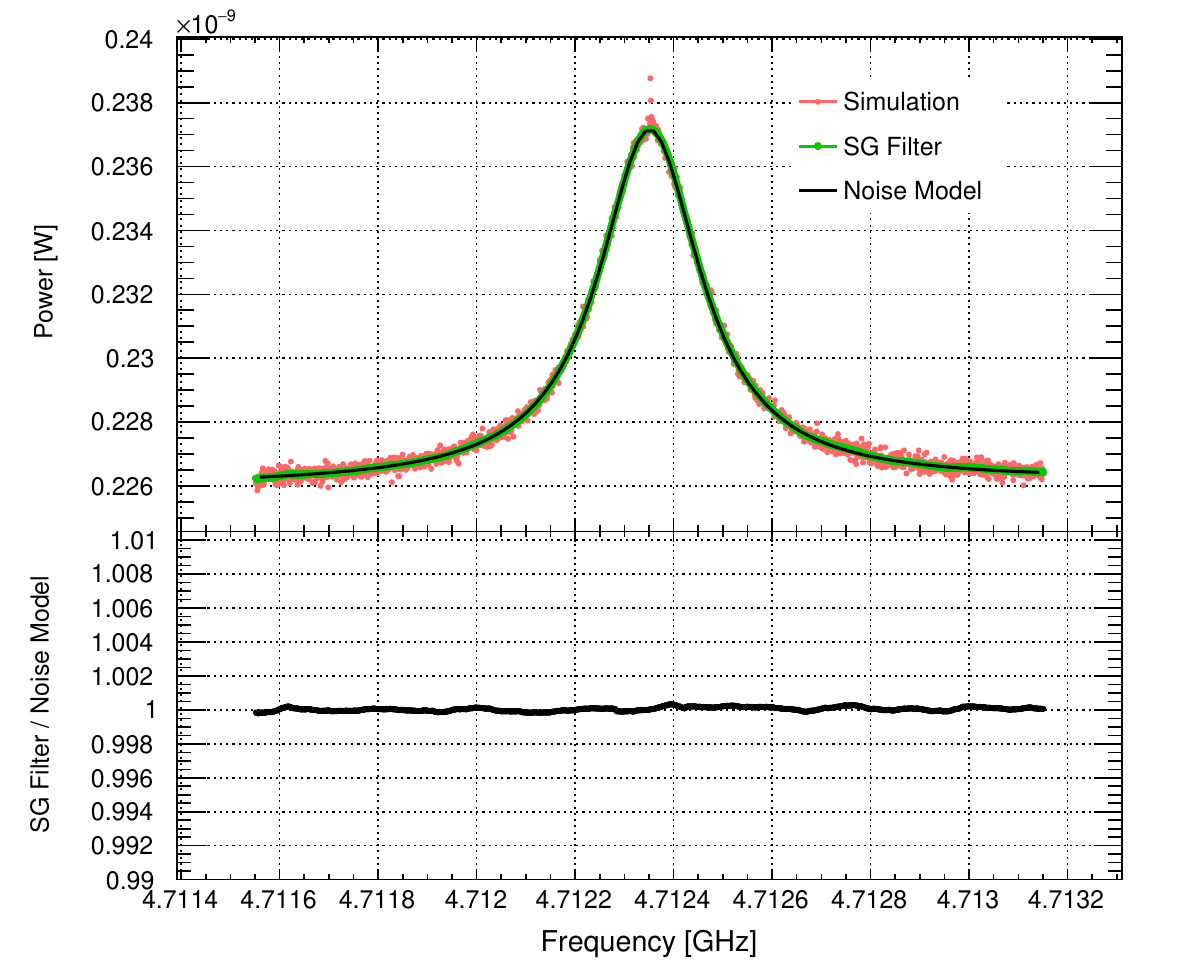}
  \caption{Upper panel: 
 The simulated spectrum (red), including the axion signal and the 
noise, is overlaid with the function that models the noise 
${\cal G}_\text{noise}$ (black) and the 
output returned by the SG filter (green). Lower panel: The ratio of the output 
returned by the SG filter to the function ${\cal G}_\text{noise}$.}
  \label{fig:sgcompare}
\end{figure}

\begin{figure} [htbp]
  \centering
  \includegraphics[width=8.6cm]{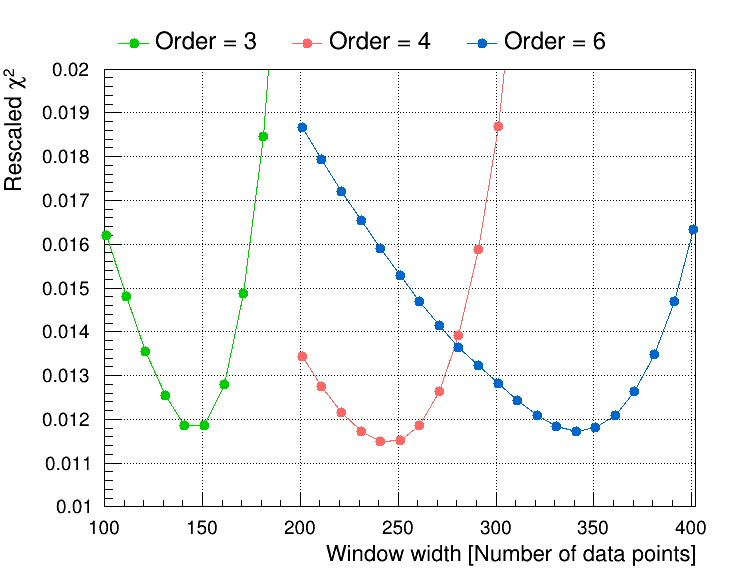}
  \caption{The rescaled $\chi^2$ when various values of window widths and 
  a 3$^\text{rd}$, a 4$^\text{th}$, or a 6$^\text{th}$-order polynomial 
  are applied in the SG filter. The rescaled $\chi^2$ is defined as 
  the sum of the squared differences  
  from all the frequency bins, between the output returned by the SG filter 
  and the function that models the noise spectrum ${\cal G}_\text{noise}$ 
  (see Fig.~\ref{fig:sgcompare}).  
  In this 
  figure, the best choice is a 4$^\text{th}$-order polynomial with 
  a window width of 241 data points (bins). }
  \label{fig:sgoptimize}
\end{figure}

\begin{figure} [htbp]
  \centering
  \includegraphics[width=8.6cm]{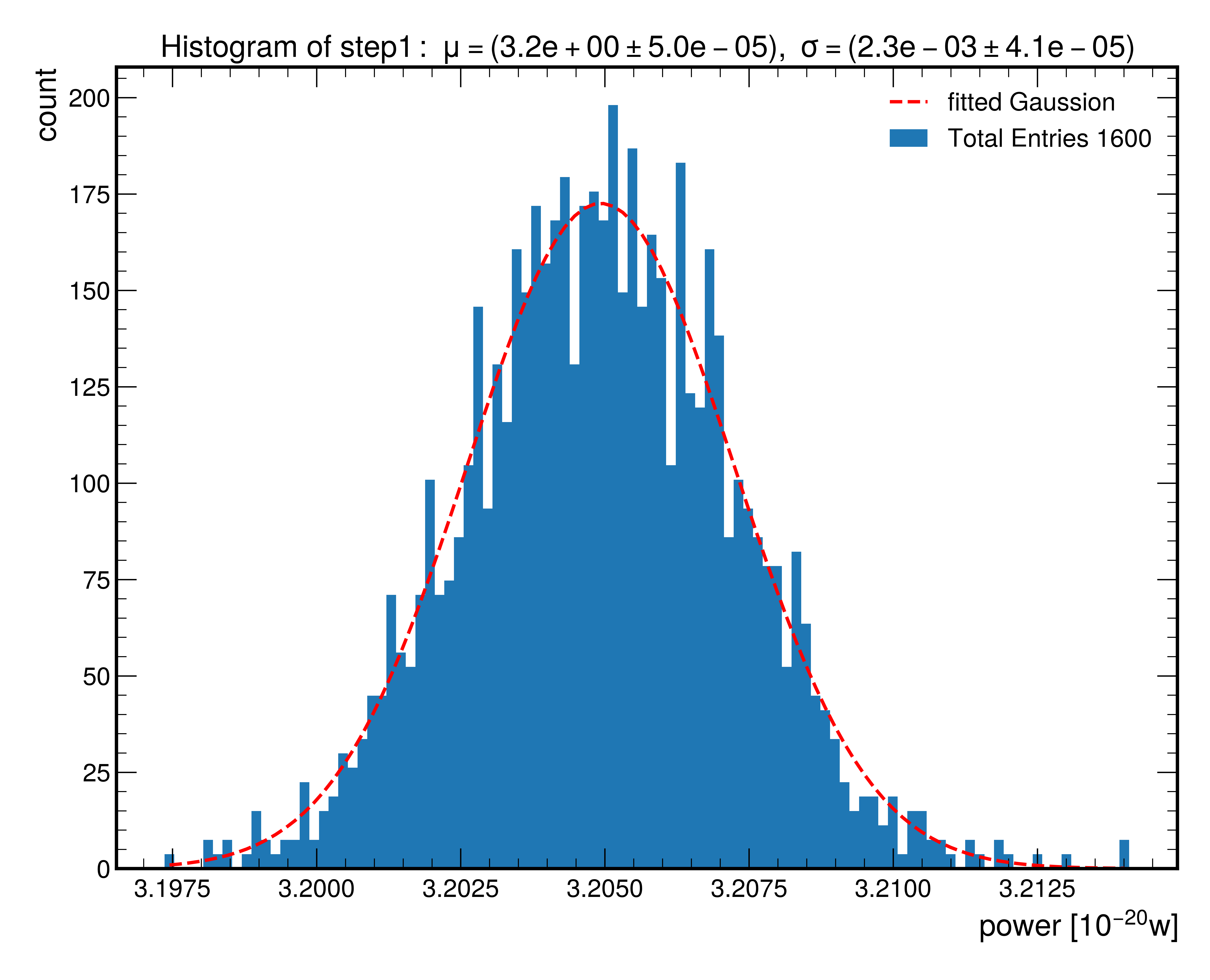}
  \caption{An example of the distribution of the measured power after 
applying the SG filter, when 
the cavity resonant frequency is 4.79815~GHz. The distribution contains 
1600 entries and each entry corresponds to the measured power 
in one frequency bin, averaged
over 1920000 subspectra. The mean and the width returned by 
a Gaussian fit to the distribution are used to determine the best choice of 
SG parameters. The fitted Gaussian mean $\mu$ divided by 
$\sqrt{1920000}$ is consistent 
with the fitted Gaussian width $\sigma$. The best choice of SG parameters 
obtained for this scan is a window of 189 data points (bins) with a 
3$^\text{rd}$-order polynomial. 
}
  \label{fig:noisegauss}
\end{figure}

The systematic uncertainties on the $\left|g_{a\gamma\gamma}\right|$ limits 
arise from the following sources:
\begin{itemize}
\item Uncertainty on the product 
$\left.Q_L\beta\middle/\left(1+\beta\right)\right.$ in Eq.~\eqref{eq:ps}: 
In order to extract the loaded quality factor $Q_L$ and the coupling 
coefficient $\beta$, a fitting of the measured results of the cavity 
scattering matrix was performed. A relative uncertainty of 3.6\% is 
assigned to this product, after a comparison of the measurements at 
$T_\text{c}\simeq155$~mK with a prediction rescaled from the measurements 
at room temperature. More details about the measurements of the 
cavity properties can be found in Ref.~\cite{TASEHInstrumentation}. 
A 3.6\% variation of this product results in a 1.9\% uncertainty 
on the $\left|g_{a\gamma\gamma}\right|$ limits. 

\item Uncertainty on the form factor $C_{010}$: 
the variation of $C_{010}$, due to the different grid sizes in the integrals 
of Eq.~\eqref{eq:formfactor}, is within 1\%, which gives a $\leq 0.5$\% 
uncertainty on the $\left|g_{a\gamma\gamma}\right|$ limits.

\item Uncertainties on the noise temperature $T_\text{a}$ from: (i) the RMS of 
the measurements in the calibration: 
$\left. \Delta T_\text{a}\middle/T_\text{a}\right.= 2.3\%$,  
and (ii) from the largest difference 
between the value determined by the calibration and that from the CD102 
data: $\left. \Delta T_\text{a}\middle/T_\text{a}\right.= 4\%$ 
(see Sec.~\ref{sec:calibration} and Fig.~\ref{fig:hemtcalvsf}). 
These two uncertainties on $T_\text{a}$ result in a 2.8\% uncertainty 
on the $\left|g_{a\gamma\gamma}\right|$ limits.

\item Uncertainty due to the misalignment (see Sec.~\ref{sec:merge}):
  estimated by comparing the central results to the one without misalignment
  ($\delta f_m = 0$)
  and to the ones with given values of $\delta f_m$.
  The comparison shows that $\delta f_m = 0$ gives the largest difference 
  of 2.8\% on the limit, which is used as the systematic uncertainty from the 
  misalignment.
  
\item Uncertainty from the choice of the SG-filter parameters: i.e.  
the window width and the order of the polynomial in the SG filter. At the 
beginning of the data taking, a preliminary optimization was performed: a 
window width of 201 bins and a 4$^\text{th}$-order polynomial were used for 
the first analysis of the CD102 data (see Sec.~\ref{sec:ana}). 
This choice is kept for the central results. 
Nevertheless, various methods of optimization are also explored. The goal 
of the optimization is to find a set of SG-filter parameters that only 
model the noise spectrum and do not remove a real signal. 
The methods include:
\begin{itemize}
 \item Minimize the difference between the two outputs returned by the SG 
filter, when the SG filter is applied to: (i) the real data only, and (ii) 
the sum of the real data and the simulated axion signals. 
 \item Minimize the difference between the output returned by the 
 SG filter and the function ${\cal G}_\text{noise}$ 
that models the noise spectrum (derived by fitting the CD102 data), 
when the SG filter is applied to the sum of the simulated noise based on 
${\cal G}_\text{noise}$ and the simulated axion signals. 
See Fig.~\ref{fig:sgcompare} for an example of the 
simulated spectrum, the function ${\cal G}_\text{noise}$, and the 
output returned by 
 the SG filter when a 3$^\text{rd}$-order polynomial and a window of 141 
 bins are chosen; the squared differences from all the frequency bins are 
summed together (rescaled $\chi^2$) when performing the optimization.
 Figure~\ref{fig:sgoptimize} shows the rescaled $\chi^2$ 
as a function of window widths when the order of polynomial is 
 set to three, four, and six. 
 \item Compare the mean $\mu_\text{noise}$ and the width $\sigma_\text{noise}$ 
of the measured power after applying the SG filter, 
assuming that no signal is present in the 
data. See Fig.~\ref{fig:noisegauss} for an example distribution 
of the measured power from the averaged spectrum of a 
single scan; a Gaussian fit is performed to extract 
$\mu_\text{noise}$ and $\sigma_\text{noise}$. Given the nature of the 
thermal noise~\cite{Dicke}, the two variables are supposed to be related to 
each other if a proper window width and a proper order are chosen:
\begin{equation*} 
\sigma_\text{noise} = \frac{\mu_\text{noise}}{\sqrt{N_\text{spectra}}},
\end{equation*}
where $N_\text{spectra}$ is the number of spectra for averaging and 
is related to the amount of integration time for each frequency step. In 
general, $N_\text{spectra}=1920000-2520000$. 
\end{itemize}

In addition, one could choose to optimize for each frequency step 
individually, optimize for a certain frequency step but apply the results to 
all data, or optimize by fitting together the spectra from all the frequency 
steps. 
The deviations from the central results using different optimization 
approaches are in general within 1\% and the 
maximum deviation of 1.8\% 
on the $\left|g_{a\gamma\gamma}\right|$ limit is used as a conservative estimate of the systematic 
uncertainty from the SG filter. 

\end{itemize}

The effects on the $\left|g_{a\gamma\gamma}\right|$ limits from these sources 
are studied and added in quadrature to obtain the total systematic uncertainty.
 The systematic uncertainties on the $\left|g_{a\gamma\gamma}\right|$ limits 
are displayed together with the central results in Sec.~\ref{sec:results}. 
Overall the total relative systematic uncertainty is $\approx 4.6\%$.

\section{Results} \label{sec:results}

\begin{figure*} [htbp]
  \centering
  \includegraphics[width=12.9cm]{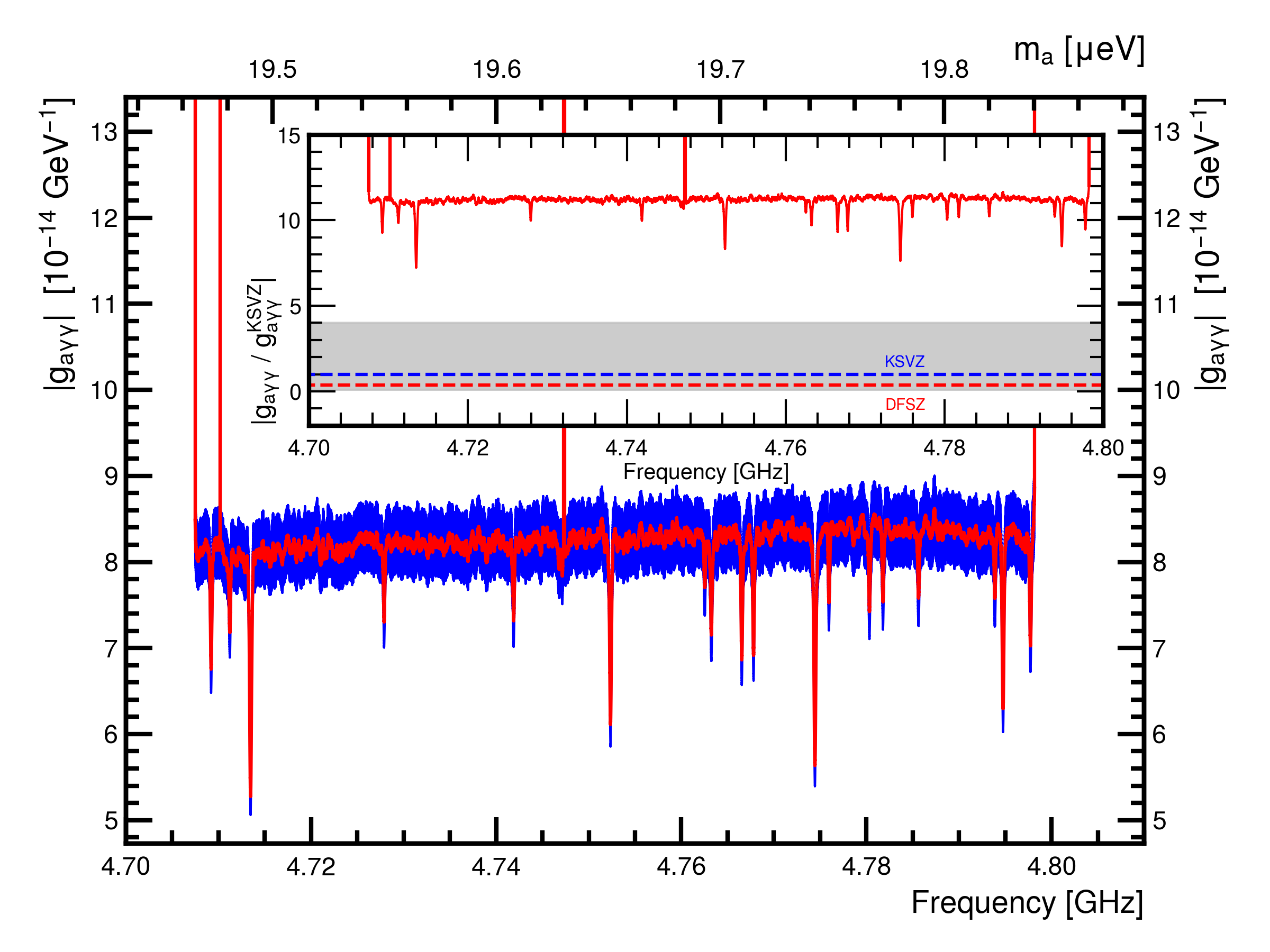}
  \caption{The 95\% C.L. limits on $\left|g_{a\gamma\gamma}\right|$ and the 
 ratio of the limits relative to the KSVZ benchmark value  
  (inset), for the frequency range of 4.70750--4.79815~GHz. The blue error 
  band indicates the systematic uncertainties as discussed in 
  Sec.~\ref{sec:sys}. The gray band in the inset shows the allowed region of 
 $\left|g_{a\gamma\gamma}\right|$ vs. $m_a$ 
 from various QCD axion models, while the blue and red dashed lines are the 
values predicted by the KSVZ and DFSZ benchmark models, respectively.}
  \label{fig:glimit}
\end{figure*}

\begin{figure*} [htbp]
  \centering
 \includegraphics[width=12.9cm]{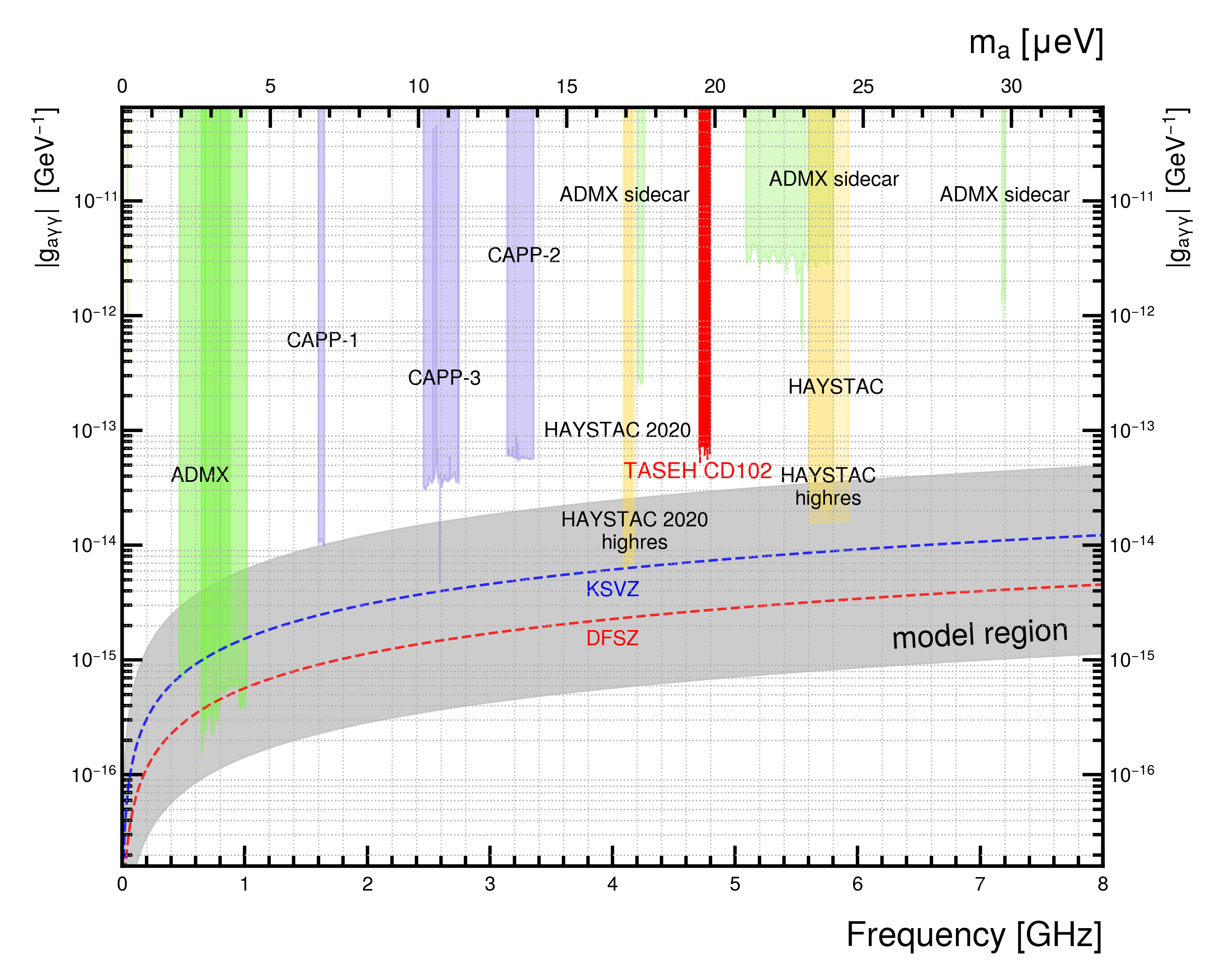}
  \caption{The limits on the axion-two-photon coupling 
 $\left|g_{a\gamma\gamma}\right|$ for the frequency range 0--8~GHz, from the 
 CD102 data of TASEH (red band) and previous searches performed by the ADMX, 
 CAPP, and HAYSTAC Collaborations. The gray band indicates the allowed region 
 of $\left|g_{a\gamma\gamma}\right|$ vs. $m_a$ from various QCD axion 
models while the blue and red dashed lines are the values predicted by the 
KSVZ and DFSZ benchmark models, respectively.}
  \label{fig:gaggall}
\end{figure*}

Figure~\ref{fig:glimit} shows the 95\% C.L. limits on 
$\left|g_{a\gamma\gamma}\right|$ and the ratio of the limits  
with respect to the KSVZ benchmark value.  
The blue error band indicates the systematic uncertainties as discussed in 
Sec.~\ref{sec:sys}. Note the uncertainties here are solely due to the 
variations in the experimental parameters and in the analysis procedure 
of TASEH. The uncertainties on the local dark matter density $\rho_a$,
 which can be as large as 50\%, are considered external uncertainties 
and not included in the blue error band.  
No limits are placed for the frequency ranges  
4.71017 -- 4.71019~GHz and 4.74730 -- 4.74738~GHz, corresponding to 
the regions in which non-axion signals were observed 
during the collection of the CD102 data. The limits on 
$\left|g_{a\gamma\gamma}\right|$ range from $5.3\times 10^{-14}\,\text{Ge\hspace{-.08em}V}^{-1}$ to $8.9\times 10^{-14}\,\text{Ge\hspace{-.08em}V}^{-1}$, 
with an average 
value of $8.2\times 10^{-14}\,\text{Ge\hspace{-.08em}V}^{-1}$; the lowest 
value comes from the frequency bins with 
additional eight times more data from the rescans, while the highest value 
comes from the frequency bins near the boundaries of the spectrum. 
Figure~\ref{fig:gaggall} displays the $\left|g_{a\gamma\gamma}\right|$ limits 
obtained by TASEH together with those from the previous searches. 
The results of TASEH exclude the models with the axion-two-photon coupling 
$\left|g_{a\gamma\gamma}\right|\gtrsim 8.2\times 10^{-14}\,\text{Ge\hspace{-.08em}V}^{-1}$, a factor of eleven above the benchmark
KSVZ model for the mass range 
$19.4687 < m_a < 19.8436 \,\mu\text{e\hspace{-.08em}V}$ (corresponding to 
the frequency range of $4.70750 < f_a < 4.79815$~GHz).

The central results in Figs.~\ref{fig:glimit}--\ref{fig:gaggall} are 
obtained assuming an axion signal line shape that follows 
Eq.~\eqref{eq:simplesignal}. The $\left|g_{a\gamma\gamma}\right|$ limits from 
the analysis that merges frequency bins without assuming a signal line shape 
are $\approx5.5$\% larger than the central values. 
If a Gaussian signal line shape with an FWHM of 2.5~kHz,  
about half of the axion line width in Eq.~\eqref{eq:simplesignal}, is 
assumed instead, the limits will be $\approx3.8$\% smaller than the central 
results. If the $\left|g_{a\gamma\gamma}\right|$ limits are derived from the 
observed SNR as described in the ADMX paper~\cite{ADMXVIII}, 
rather than using the 5$\sigma$ target SNR, the average limit on 
$\left|g_{a\gamma\gamma}\right|$ will 
be $\approx 4.9\times 10^{-14}\,\text{Ge\hspace{-.08em}V}^{-1}$.

\section{Conclusion} \label{sec:conclusion}
This paper presents the analysis details of a search for axions for the mass 
range $19.4687 < m_a < 19.8436 \,\mu\text{e\hspace{-.08em}V}$, using the 
CD102 data collected by the Taiwan Axion Search Experiment with Haloscope 
from October 13, 2021 to November 15, 2021. 
Apart from the non-axion signals, no candidates with a significance more than
3.355 were found. The synthetic 
axion signals were injected after the collection of data and the 
successful results validate the data acquisition and the analysis procedure. 

The experiment excludes models with the 
axion-two-photon coupling $\left|g_{a\gamma\gamma}\right|\gtrsim 8.2\times 10^{-14}\,\text{Ge\hspace{-.08em}V}^{-1}$ at 95\% C.L.,
 a factor of eleven above the benchmark KSVZ model. The sensitivity on 
$\left|g_{a\gamma\gamma}\right|$ reached by TASEH 
is three orders of magnitude better than the existing limits in the same 
mass range.  
It is also the first time that a haloscope-type experiment places 
constraints in this mass region. The readers shall be aware that haloscope 
experiments assume that 100\% of the dark matter is the axion. In addition, 
the local dark matter density, which is used to compute the expected axion 
signal power, can have an uncertainty as large as 50\%; this uncertainty 
is typically considered an external 
uncertainty and not included in the experimental results. 

The target of TASEH is to search for axions for the mass range of 
16.5--20.7$\,\mu\text{e\hspace{-.08em}V}$ corresponding to a frequency range 
of 4--5~GHz, with a capability to be extended to 2.5--6~GHz in the future. 
In the coming years, several upgrades are expected, including: the use of a 
quantum-limited Josephson parametric amplifier as the first-stage amplifier, 
the replacement of the existing dilution refrigerator with a new one that has 
a magnetic field of about 9~Tesla and a larger bore size, and the development 
of a new cavity with a significantly larger effective volume. 
With the improvements of the experimental setup and several years of data 
taking, TASEH is expected to probe the QCD axion band in the target mass range.

\begin{acknowledgments}
We thank Chao-Lin Kuo for his help to initiate this project as well as 
discussions on the microwave cavity design, Gray Rybka and Nicole Crisosto 
for their introduction of the ADMX experimental 
setup and analysis, Anson Hook for the discussions and the review of the 
axion theory, and Jiunn-Wei Chen, Cheng-Wei Chiang, Cheng-Pang Liu, and 
Asuka Ito for the discussions of future improvements in axion searches.  
  The work of the TASEH Collaboration was funded by 
the Ministry of Science and Technology (MoST) of Taiwan with grant numbers 
MoST-109-2123-M-001-002, MoST-110-2123-M-001-006, MoST-110-2112-M-213-018, 
MoST-110-2628-M-008-003-MY3, 
and MoST-109-2112-M-008-013-MY3, and by the Institute of Physics, Academia 
Sinica. 

\end{acknowledgments}

\appendix
\section{Derivation of the Function that Models the Noise Spectrum} 
\label{sec:cavitynoise}

The background noise from a cavity is governed by the thermal noise and the 
vacuum fluctuation. 
According to Planck's law in one dimension (1D), the spectral density of 
the electromagnetic noise from the cavity, thermalized with an environment of 
temperature $T_{\rm c}$, through a transmission line is 
\begin{equation}
\label{eq:cavity_thermal_spectral_density}
    S(\omega) = \hbar\omega \left( \frac{1}{e^{\hbar\omega/k_{\rm B}T_{\rm c}} - 1} +\frac{1}{2} \right),
\end{equation}
where $\omega$ is the angular frequency, $\hbar$ is the reduced Planck's 
constant, and $k_{\rm B}$ is the Boltzmann constant.

\begin{figure}
    \centering
    \includegraphics[width=8.6cm]{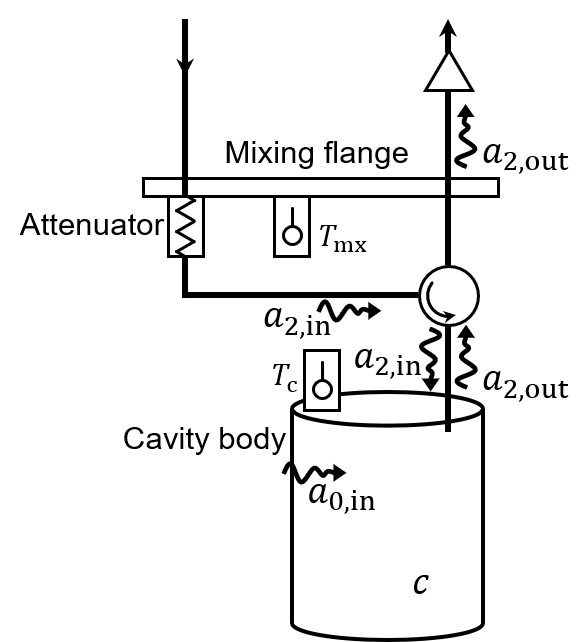}
    \caption{The input-output model of cavities. The cavity field mode $c$ is 
coupled to the modes of a 1D transmission line $a_2$ (with a rate $\kappa_2$)
 and the modes of the cavity body $a_0$ (with a rate $\kappa_0$). 
The incoming and outgoing fields of the transmission line 
are separated by the circulator. The attenuator and the cavity body emitting 
the fields $a_{2,\rm in}$ and $a_{0,\rm in}$ are thermalized at $T_{\rm mx}$ 
and $T_{\rm c}$, respectively. 
}
    \label{fig:cavity_in_out}
\end{figure}

However, the cavity body (the materials that form the cavity itself) may not 
be thermalized with its 1D electromagnetic environment. To understand the 
noise spectrum from the cavity near its 
resonant frequency $\omega_{\rm c}/2\pi$ in this scenario, the model in 
Fig.~\ref{fig:cavity_in_out} is considered. Through a probe the cavity field 
mode $c$ is coupled to the modes $a_2$ of a 1D transmission line, representing
 the path toward a signal receiver, with a rate $\kappa_2$. The cavity field 
is also coupled to the modes of the cavity body $a_0$, representing the 
intrinsic loss, with a rate $\kappa_0$. In a steady state, 
the quantum input-output theory leads to a relation between the outgoing field 
from the cavity to the 1D transmission line, $a_{\rm 2,out}$, and the incoming
 fields, $a_{\rm 2,in}$ and $a_{\rm 0,in}$, through the elements of the 
cavity scattering matrix:
\begin{equation} \label{eq:outgoing_field}
	 a_{\rm 2,out} = S^{*}_{\rm 22} a_{\rm 2,in} + S^{*}_{\rm 20} a_{\rm 0,in},
\end{equation}
where $S_{\rm 22} = \frac {\kappa_0 - \kappa_{\rm 2} + i 2 \Delta} {\kappa_0+\kappa_{\rm 2} + i 2 \Delta}$, $S_{\rm 20} = \frac {2\sqrt{\kappa_0 \kappa_{\rm 2}}} {\kappa_0+\kappa_{\rm 2} + i 2 \Delta}$,
and $\Delta = \omega-\omega_{\rm c}$ is the detuning. 

As both incoming fields are in a thermal state, $\langle a_{i,\rm in}(0) a_{i,\rm in}(\tau) \rangle = n_{\rm th}(T_i) \delta(\tau)$, where $n_{\rm th}(T_i) = \frac{1}{e^{\hbar\omega/k_{\rm B}T_i} - 1}$ is the mean thermal photon number 
of the incoming field $a_{i,\rm in}$ at the temperature $T_i$, and 
$\delta(\tau)$ is the $\delta$-function. In the model the incoming field 
$a_{\rm 2,in}$ comes from a nearby attenuator, anchored to the mixing flange, 
in the transmission line 
with a temperature $T_{\text mx} \equiv T_{\rm 2}$, and $a_{\rm 0,in}$ comes 
from the cavity body with a temperature $T_{\text c} \equiv T_{\rm 0}$.

By defining the effective temperature $\Tilde{T}_i = \left( n_{\rm th}(T_i) + \frac{1}{2} \right) \hbar\omega/k_{\rm B}$,
the power spectral density of the outgoing field of the transmission line 
modes is
\begin{equation}
\label{eq:spectrum_relation}
\begin{aligned}
    S_{\rm out}(\omega) 
    =& \int_{-\infty}^{\infty} \hbar\omega \left(\langle a_{2,\rm out}(0) a_{2,\rm out}(\tau) \rangle + \frac{1}{2} \right) e^{-i\omega\tau} d\tau \\
    =& |S_{\rm 22}|^2 k_{\rm B} \Tilde{T}_2 + |S_{\rm 20}|^2 k_{\rm B} \Tilde{T}_0.
\end{aligned}
\end{equation}
The total output noise can be viewed as the sum of the reflection of the 
incoming noise from the attenuator and the transmission of the noise from the 
cavity body itself.
Via the unitary property of 
the cavity scattering matrix, i.e. $|S_{\rm 22}|^2+|S_{\rm 20}|^2=1$,
\begin{equation}
\label{eq:spectrum_relation_Lorentz}
\begin{aligned}
    S_{\rm out}(\omega) =  k_{\rm B} \Tilde{T}_2 + k_{\rm B} (\Tilde{T}_0 - \Tilde{T}_2) L(\omega),
\end{aligned}
\end{equation}
where $L(\omega) = |S_{\rm 20}|^2 = \frac {\kappa_0 \kappa_{\rm 2}}  {(\kappa_0+\kappa_{\rm 2})^2/4 + \Delta^2}$ is a Lorentzian function with a FWHM 
$\kappa_0+\kappa_{\rm 2}$. Therefore, the noise spectrum has a flat background
 determined by the incoming noise of the attenuator with an effective 
temperature $\Tilde{T}_{\rm 2}$, plus an excess Lorentzian peak centered at 
$\omega_{\rm c}$ determined by the effective temperature difference 
$\Tilde{T}_{\rm 0} - \Tilde{T}_{\rm 2}$. (The center Lorentzian structure 
can even be a dip if ${T}_{\rm 0} < {T}_{\rm 2}$.)

\providecommand{\noopsort}[1]{}\providecommand{\singleletter}[1]{#1}%

\end{document}